\documentclass[useAMS,usenatbib]{mn2e}

\usepackage{graphicx}
\usepackage{bm}

\title[Hybrid AGB winds]{A hybrid steady-state magnetohydrodynamic dust-driven stellar wind model for AGB stars}
\author[Anand Thirumalai and Jeremy S. Heyl]{Anand Thirumalai$^{1}$\thanks{E-mail:anand@phas.ubc.ca (AT); heyl@phas.ubc.ca (JSH)} and Jeremy S. Heyl$^{1}$\footnotemark[1]\\
$^{1}$University of British Columbia, 6224 Agricultural Road, Vancovuer, British Columbia, V6T 1Z1, Canada}

\begin{document}

\date{\today}

\pagerange{\pageref{firstpage}--\pageref{lastpage}} \pubyear{2010}

\maketitle

\label{firstpage}

\def\aj{AJ}                   
\def\apj{ApJ}                 
\def\apjl{ApJ}                
\def\aap{Astron. \&  Astroph.}                
\def\araa{ARA\&A}             
\def\apjs{ApJ}                
\def\apss{ApSS}		
\def\mnras{MNRAS}             
\def\nat{Nature}              
\def\physrep{Phys.~Rep.}      
\def\pra{Phys. Rev. A} 			
\def\pre{Phys. Rev. E} 			
\def\prb{Phys. Rev. B} 			
\def\prd{Phys. Rev. D} 			
\def\pasp{Pubs. Astron. Soc. Pac.}
\def\solphys{Sol. Phys.}

\begin{abstract}
  We present calculations for a magnetised hybrid wind model for
  Asymptotic Giant Branch (AGB) stars. The model incorporates a
  canonical Weber-Davis (WD) stellar wind with dust grains in the
  envelope of an AGB star. The resulting hybrid picture preserves
  traits of both types of winds. It is seen that this combination
  requires that the dust-parameter ($\Gamma_{d}$) be less than unity
  in order to achieve an outflow. The emergence of critical points in
  the wind changes the nature of the dust-driven outflow,
  simultaneously, the presence of a dust condensation radius changes
  the morphology of the magnetohydrodynamic (MHD) solutions for the
  wind. In this context, we additionally investigate the effect of
  having magnetic-cold spots on the equator of an AGB star and its
  implications for dust formation; which are seen to be consistent
  with previous findings.
\end{abstract}

\begin{keywords}
AGB wind, Weber-Davis wind, MHD,  dust-driven wind.
\end{keywords}

\section{Introduction}\label{sec:intro}

In recent years several observations of evolved stars and planetary
nebulae have indicated that magnetic fields may exist in these objects
(e.g. \citealt{Amiri2010}, \citealt{Jordan2005}, \citealt{Herpin2006},
\citealt{Sabin2007_1} \citealt{Sabin2007_2} and
\citealt{Miranda2001}). The inferred magnetic fields in such objects,
at the distances of the masers, indicate a variety of field strengths
ranging from a few milligauss to a hundred gauss or so
\citep{Vlemmings2005, Vlemmings2006}. This in turn, may indicate a variety of field
strengths at the surfaces of such objects. The idea of studying the
effects of a magnetic field on the nature of the AGB wind has been
carried out mainly with regard to explaining the diversity of the
observed shapes (see for example \citealt{Chu1987} and
\citealt{Stanghellini1993} and references therein) of planetary
nebulae. As a result, the modelling of magnetic fields in these stars
has focussed on the final stages of the AGB phase, at the very tip of
the AGB, or on the post-AGB phase itself, employing the so called
interacting wind scenario \citep[see][for outlines of the key points of
interacting wind
models]{Frank1999,Franco2001,Gardiner2001}.

It has therefore been argued that magnetic fields play a vital role in
shaping planetary nebulae by several researchers. However, it is not completely certain whether
magnetic fields play a dynamic global or local role in this regard. The reader is referred to work carried out by Soker and co-workers \citep[e.g.][]{Soker2003,Soker2002_1,Soker2002_4,Soker1999,Soker1999_2,Soker1998} for discussion regarding this topic. Moreover, current state-of-the-art MHD models for the winds of stars at the tip of the AGB, do not take into account the effect of
radiation pressure on the dust grains in the envelope of an AGB
star. It is generally thought that the mass loss in AGB stars is
largely governed by this mechanism coupled with strong stellar
pulsations
\citep{Lamers,Elitzur2003,Bowen1988,Bowen1991,Fleischer1992,Wood1979,Bedijn1988}. However,
to the best of our knowledge, there have not been any investigations
combining a standard dust-driven wind scenario with MHD effects, in the
literature. Such a study, given the importance of magnetic activity in
AGB stars, would help bring together two different sub-classes of
stellar winds. This is the aim of the current paper; to investigate
the implications of combining magneto-rotational effects with a
dust-driven wind in AGB stars. It is to be mentioned that this would
be applicable at the early stages of the AGB phase, long before the
interacting wind scenario becomes important, wherein the models
mentioned above would be more likely candidates for describing the
outflows.

At this juncture, we conduct a brief survey of the available
literature with regard to both dust-driven winds and
magneto-rotational equatorial winds. However, the reader is referred
to a review by \citet[][and references therein]{Tsinganos2007} for MHD
outflows and likewise, reviews by \citet{Dorschner2003} and
\citet{Habing2004} for historical reviews of AGB stars and dust-driven
winds. Thereafter, the model developed in the current study, is
elucidated.

With regard to magneto-rotational equatorial winds, seminal work was
carried out by \citet[][WD]{WD67}. They formulated a steady state
description for the radial and azimuthal components of the solar
wind's momentum. Essentially the same results were also arrived at
independently, by \citet{Mestel1967}. These studies formed the
groundwork upon which further studies were conducted. Thereafter,
magnetic braking by a stellar wind was also investigated by
\citet{Mestel1968} and later in greater detail by
\citet{Okamoto1974,Okamoto1975}, wherein the theory was extended to
cover a variety of field configurations with poloidal fields. The
Weber and Davis equatorial wind theory was extended by
\citet{Goldreich1970} towards a relativistic treatment of the wind,
including the effects of pressure and gravitation.  \citet{Michel1969}
carried out a similar analysis of the WD model but neglected pressure
and gravity and thus, relativistic magnetosonic critical points do not
appear in his model. The first effort to investigate the importance of
the interactions between the gas and the magnetic field, with regard
to determining properties of the structure and dynamics of the solar
corona, was carried out by \citet{Pneuman1971}. They found that the
assumed dipolar field configuration had a profound effect on the solar
wind in creating streamers. \citet{Yeh1976} conducted a parametric
study of the mass and angular momentum effluxes of magneto-rotational
stellar winds and found that the mass efflux would be large, if the
mass of the star was small, with a large radius, provided the stellar
corona was dense and hot. Simultaneously, he found that the angular
momentum efflux became greater when the magnetic field and stellar
rotation parameters were increased. \citet{Belcher1976} revisited the
Weber and Davis theory and identified two regimes; the so-called slow
and fast magnetic rotators, which are defined by the ratios of the
Michel and Parker velocities of the wind. The former is related to the ratio of the magneto-rotational flux to the mass flux and the latter is related to the squares of the sound speed, escape velocity and radial bulk gas velocity at the surface. Presently, the reader is referred to \citet{Belcher1976} for mathematical expressions for these quantities. Belcher and MacGregor were able to delineate the angular
momentum evolution of the two types of rotators on the main
sequence. \citet{Barker1982} extended the Weber-Davis
theory for the case of non-zero photospheric mass loss, showing that
including a dimensionless constant corrected the original theory. The
original WD theory was also re-analysed elsewhere, with an effort to
study standing MHD shocks by \citet{Chakrabarti1990}. Various
solutions of the WD solution topology were studied and he found that
many of them allowed for MHD shock formation, in both accretion and
winds, of a compact magnetised object. With the advent of increased
computational capabilities of computers in the 1980's,
\citet{Sakurai1985} successfully generalised the original Weber and
Davis theory to two dimensions. There appear the usual slow and fast
modes in his solution, as in the original Weber and Davis theory;
however, in this case, the momentum equation was found to be singular
on an Alfv\'en surface and regularising the solution on this surface
alongside the boundary condition at the photosphere uniquely
determined the solution of the two-dimensional
problem. \citet{Keppens1999,Keppens2000} have over the past few years,
developed two- and three-dimensional MHD models for
investigating magento-rotational stellar winds. Their models have
clearly shown that the dipolar nature of the magnetic field structure
is important for stellar winds.  At the same time they have shown that
the poloidal component leads to density enhancement along the
equatorial region. They also find that their trans-sonic wind
solutions have dead zones which have a latitudinal
dependence; this can be traced back to the configuration of the
magnetic field. They have also investigated shock formation in
magneto-rotational outflows \citep[see]{Keppens2003} and the nature
and formation of Kelvin-Helmholtz instabilities
\citep[e.g.][]{Keppens1999_2}. These studies indicate that the mass
efflux from magneto-rotational outflows can be asymmetric in
nature. Their models also have been used to study the evolution of
rotational velocity distribution in late-type stars while on the main
sequence. The stellar winds in these stars were assumed to be WD-like
steady-state winds \citep[see][]{Keppens1995}.

Concomitant to the development of MHD winds, the field of dust-driven
winds in evolved stars has flourished as a separate research focus
altogether. The early models of dust-driven outflows from AGB stars
were one-dimensional and did not include radiative transfer
through the dust-laden envelope and dynamical dust formation
\citep[see][for a brief review]{Lamers}. Later one-dimensional models
involved dynamics of dust formation and growth in a time-dependent
manner building upon initial studies of C-type AGB stars
\citep[e.g.][]{Gail1988,Gauger1990,Fleischer1991}. These models
revealed, that due to the dynamical interaction between dust and gas
and the radiation field, dust-laden shells were formed in the
envelopes of AGB stars \citep[e.g][]{Fleischer1992} and dynamical
instabilities in the flow also manifested themselves leading to
aspherical mass loss
\citep[see][]{Fleischer1995,Hofner1995,Hofner1996}.  Models have also
been constructed that investigate dust grain drift through the gas
which modifies dust growth rates and the efficiency of the wind
acceleration process \citep[see][]{Sandin2004}. There have also been
further improvements in treating the radiative transfer in a frequency
dependent way alongside detailed micro-physics of molecular and grain
opacities \citep[e.g.][]{Hofner2003,Hofner2003_2}. The reader is
presently referred to a recent review by \citet{Hofner2008} on the
status of modern-day radiation-hydrodynamical modelling of dust-driven
winds of AGB stars. The modelling of dust-driven winds has
elsewhere, been extended to more than one dimension, capturing the aspherical
nature of the outflow in a very clear way
\citep[see][]{Woitke2005,Woitke2005_2,Woitke2006,Woitke2008_2,Woitke2008}. These
models incorporate hydrodynamics with radiation pressure on the dust,
equilibrium chemistry for the nucleation of dust in a time-dependent
way, and they take into account radiative transfer in a frequency
dependent manner. They were able to capture in their simulations, many
of the highly dynamical aspects of the outflow, including turbulent
and inhomogeneous dust formation and Rayleigh-Taylor flow
instabilities that result in cloud-like structure formations in the
efflux. With the help of such models it has become clear that the
winds in these stars are far more complicated than simple
one-dimensional (spherically symmetric) pictures and may have an
impact on shaping planetary nebulae due to their asphericity
\citep[e.g.][]{Hofner2000} or by having an impact on the superwind at
the end of the AGB, as is suggested by \citet{Lagadec2008}, with regard
to the abundance of carbon in the envelopes of some AGB stars. The
latter suggests that the effort to include complicated micro-chemistry
of the dust grains, as is done in modern AGB wind models, may prove to
be a key factor in resolving issues regarding the superwind in the
late stages of the AGB phase. For a more thorough description of AGB
envelopes and the stochastic nature of dust formation process itself,
the reader is referred to \citet{Habing2004} and to \citet{Dirks2008}
as well as references therein.

It has also been argued that asymmetric mass loss on the AGB may
result in kicks and spin being delivered to the nascent white dwarf
within \citep[see][]{Spruit1998}. More recently, the effect of kicks
to white dwarfs has been investigated with regard to their impact on
the dynamics of globular clusters and binaries
\citep[see][]{2008MNRAS.383L..20D,Heyl2007,Heyl2007_2,Heyl2008_2,Heyl2008,Heyl2009}. These
studies indicate that the study of AGB winds in relation to
asphericity is rather important. In the literature, hybrid equatorial
winds have only been considered for rotating hot stars
\citep[see][]{Friend1992,Friend1987,Friend1984,Friend1989,Friend1986}.
MHD stellar winds for AGB stars have been simulated for a coupled disk
and star system, clearly showing that magnetocentrifugal winds can
occur in planetary nebulae and AGB stars, at the tip of the AGB
\citep[see][and references therein]{Blackman2001}. Other researchers
have considered a hybrid wind for AGB stars by coupling Alfv\'en waves
with radiation pressure \citep[see][]{Falceta2002,Falceta2006}, showing
that it is possible to obtain low-velocity mass efflux in supergiant
cool stars through such a mechanism. A similar study has also been
carried out for Wolf-Rayet stars \citep[e.g][]{Dossantos1993}, of
course obtaining much faster winds. From the brief survey of the
literature conducted above, it is evident that winds from AGB stars
whether during the initial or final phases of the AGB can have
asymmetries that result in aspherical mass loss either due to purely
MHD effects, or due to the dynamics between the dust and gas and the
radiation field. However, present day dust-driven models still do not
include magneto-rotational effects, particularly when more and more
observations indicate the presence of magnetic fields in these
objects and the pure MHD wind models do not include the effect of
dust condensation and radiation pressure in the envelopes of these
stars. The aim of the current work is to investigate the implications
of including magneto-rotational effects with a simplified dust-driven
model in 1.5 dimensions; the azimuthal terms are determined entirely from their dependence on the radial terms (the standard WD-picture). We shall however, for the sake of simplicity, be
considering a steady-state case. It is to be re-iterated 
that the current model may only be valid at the early phases
of the AGB, long before the onset of the superwind at the end of the
AGB.

We present the underlying assumptions of the model in the following
section (\S\ref{sec:model}). In \S\ref{sec:numer} we present the
numerical details, thereafter in \S\ref{sec:Results and Discussion}
the results are presented and discussed. Finally, in
\S\ref{sec:Conclusion} the paper is summarised and future avenues for
development of the current model are briefly discussed.

\section{The hybrid wind model}\label{sec:model}

We begin with the standard picture of a WD-equatorial wind; we shall
be following their definitions and general derivation
closely. However, we shall be including the dust as a second fluid. As
in the WD model we shall assume complete axial symmetry and the
customary explicit form for the magnetic field within the equatorial
plane,
\begin{equation}
\vec{B}=B_r(r)\hat{r}+B_{\phi}(r)\hat{\phi},
\label{eq:1}
\end{equation}
while the velocity of the gas and the dust can respectively, be written as,
\begin{equation}
\vec{u}=u_r(r)\hat{r}+u_{\phi}(r)\hat{\phi}
\label{eq:2}
\end{equation}
and
\begin{equation}
\vec{v}=v_r(r)\hat{r}+v_{\phi}(r)\hat{\phi}.
\label{eq:3}
\end{equation}
As can be seen, the velocity fields are functions of the radial
distance alone. In addition, there is no time dependence, explicit or
implicit, thus conforming with the steady-state
assumption. Accordingly, the continuity equation for the gas and dust
combined can be written as,
\begin{equation}
\rho u r^2 + n_d m_d v r^2 \Theta(r-r_d) = \mathrm{constant},
\label{eq:4}
\end{equation}
where, $\rho$ is the gas density in cgs units, $n_d$ is the number
density of the dust grains and $m_d$ is the mass of a single dust
grain. $\Theta(r-r_d)$ is the standard Heaviside function which is
equal to unity for $r \geq r_d$. Prior to the dust formation radius,
there is only one fluid, namely the gas. Please note that we have
dropped the subscript $r$ from the radial velocities for brevity and
future convenience. We are explicitly assuming, for the sake of
simplicity, that all dust formation occurs at a certain distance
($r_d$) from the centre of the star. Beyond this, there is no further
condensation of dust. We shall also assume that all the dust grains
are identical and perfectly spherical, with a radius of $a=0.05 \mu$m
and a mass density of $\rho_d \approx 2.25$g/cm$^3$
\citep[e.g.][]{Lamers}. This yields the mass for an individual dust
grain; $m_d=\frac{4}{3} \pi a^3 \rho_d$.

In addition, we shall assume that the dust-to-gas ratio in the stellar
wind is given by,
\begin{equation}
\frac{n_d m_d}{\rho}=\delta \leq \frac{1}{200},
\label{eq:5}
\end{equation}
following \citet{Lamers}, in other words, $n_d m_d \ll \rho$. The maximum value for this ratio is at the dust formation radius and it decreases monotonically thereafter. For
deriving the momentum equations for the gas and the dust, our starting
point is the Euler equation, one for each of the fluids. For the gas
we can write,
\begin{equation}
(\vec{u} \cdot \vec{\nabla})\vec{u} + \frac{1}{\rho} \vec{\nabla}p + 
\frac{GM_{*}}{r^2} \hat{r} - \frac{1}{\rho c} \vec{J} \times \vec{B} - \frac{n_d}{\rho}\vec{f}_{D}=0,
\label{eq:6}
\end{equation}
where $p$ is the gas pressure and $\vec{J}$ is the current density. It
is assumed here implicitly that there is no relative motion of the
ions with respect to the neutrals. The first and second terms are
related to the velocity and pressure gradients respectively, the third
term is the gravitational acceleration on the gas, the fourth term
is the Lorentz force divided by the gas density and the final term is
proportional to the drag force that is experienced by the gas, due to
the dust grains moving through it. Assuming force-free MHD it can be shown that \citep[e.g.][]{WD67}
\begin{equation}
r(uB_{\phi}-u_{\phi}B_r)=\mathrm{constant}=-R_0^2\Omega B_{r,0},
\label{eq:7}
\end{equation}
where, $R_0$ is the stellar radius (radius of the photosphere),
$\Omega$ is the rotation rate of the star and $B_{r,0}$ is the radial
component of the magnetic field at the stellar surface.Requiring that
$\vec{\nabla} \cdot \vec{B}=0$, yields the familiar relation
\begin{equation}
r^2B_r=R_0^2B_{r,0}.
\label{eq:8}
\end{equation}
Presently, let us turn our attention to the Euler equation for the
second fluid, viz., the dust grains. The rationale is that
obtaining expressions for the drag force components will allow us to
re-write Eq.~(\ref{eq:6}), the Euler equation for the gas. The Euler
equation for the dust grains can be written as,
\begin{equation}
(\vec{v} \cdot \vec{\nabla})\vec{v} + \frac{GM_{*}}{r^2} \hat{r} - \frac{\pi a^2 Q_{rp} L_{*}}{4 \pi r^2 c m_d} \hat{r} + \frac{1}{m_d}\vec{f}_{D}=0.
\label{eq:9}
\end{equation}
In the above equation, $Q_{rp}$ is the radiation pressure mean
efficiency \citep[e.g.][]{Lamers}. In the current study, since our aim
is to present a simplistic picture we shall not be calculating this
term. $L_*$ is the luminosity of the star and $c$ is the speed of
light. The first term in Eq.~(\ref{eq:9}) is related to the velocity
gradient of the dust grains, the
second term is the gravitational acceleration experienced by the dust
grain, the third term is the radiation pressure that the dust grain
experiences, that drives it outward and the final term is the drag
force per unit mass of the dust grain, as it moves through the
surrounding gas. At this stage, we can make the usual simplifying
assumption that for a single dust grain, the radiation pressure and
the drag force terms in Eq.~(\ref{eq:9}) dominate completely over the
other terms \citep[e.g.][]{Lamers} and balance each other. We can
then write,
\begin{equation}
\frac{\pi a^2 Q_{rp} L_{*}}{4 \pi r^2 c m_d} \hat{r} - \frac{1}{m_d}\vec{f}_{D}=0.
\label{eq:10}
\end{equation}
We can now write the radial and azimuthal components of
Eq.~(\ref{eq:10}) separately. Each of these components must
identically vanish. Thus we get for the radial component,
\begin{equation}
\frac{\pi a^2 Q_{rp} L_{*}}{4 \pi r^2 c} = f_{D}^r,
\label{eq:11}
\end{equation}
where $f_{D}^r$ is the radial component of the drag force, defined as,
\begin{equation}
f_D^r=\pi a^2 \rho (v-u)\sqrt{(v-u)^2 + a_{th}^2}~,
\label{eq:12}
\end{equation}
where $a_{th}$ is the thermal speed given by $a_{th}=\sqrt{2kT/\mu
  m_u}$ and $\mu m_u$ is the mean molecular mass of the
gas. Typically for AGB stars with nearly solar abundance with
pulsational shocks that extend the density structure we can use $\mu
\approx 1.3$ \citep[e.g.][]{Bowen1988}.  From Eq.~(\ref{eq:10}) we can
immediately write the azimuthal momentum equation for the dust grains
as,
\begin{equation}
\frac{f_{D}^{\phi}}{m_d} = 0,
\label{eq:13}
\end{equation}
where $f_{D}^{\phi}$, the azimuthal component of the drag force,
implying that there is no drag in the azimuthal direction.  That is,
the dust is co-rotating with the gas. Having obtained
expressions for the radial and azimuthal components for the dust
grains we can now re-visit the Euler equation for the gas. The radial
momentum equation for the gas can be re-cast in the form,
\begin{equation}
u\frac{du}{dr}-\frac{u_{\phi}^2}{r} + \frac{1}{\rho}\frac{dp}{dr} + \frac{GM_{*}(1-\Gamma_d)}{r^2} - \frac{1}{\rho c} (\vec{J} \times \vec{B})_r =0,
\label{eq:14}
\end{equation}
wherein, we have the usual definition that,
\begin{equation}
\Gamma_d = \frac{n_d}{\rho} \frac{\pi a^2 Q_{rp} L_{*}}{4 \pi c G M_{*}}.
\label{eq:15}
\end{equation}
Presently, the azimuthal momentum equation for the gas can be written, after some re-arrangement, as,
\begin{equation}
\rho u r^2 \frac{d}{dr}(r u_{\phi}) - r^2 B_r \frac{d}{dr}(r B_{\phi})=0.
\label{eq:16}
\end{equation}
It is to be noted that the azimuthal component of the drag force does
not appear in Eq.~(\ref{eq:16}) as it vanishes (see
Eq.~(\ref{eq:13})). Moreover, since the dust-to-gas ratio is small, i.e., $n_d m_d \ll
\rho$ and since the dust and gas velocities are expected to be on the
same order with the dust velocity exceeding the gas velocity by a
reasonable fraction of the gas velocity, it is reasonable to make the
approximation that $\rho u r^2 \gg n_d m_d v r^2$. This allows us to
further make the approximation that $\rho u r^2 = \mathrm{constant}$ in
Eq.~(\ref{eq:4}). Therefore, we can immediately recover the usual expression for the total
specific angular momentum of the wind as,
\begin{equation}
r u_{\phi} - \frac{B_r}{4 \pi \rho u} r B_{\phi} = L = \Omega r_A^2,
\label{eq:17}
\end{equation}
where $r_A$ is the Alfv\'{e}n radius, defined as the distance from the
centre of the star at which the radial magnetic energy density is equal to the
kinetic energy density, i.e., $\frac{1}{2} \rho u^2 = \frac{B_r^2}{8\pi}$. At this stage, we assume a polytropic equation of state for the gas as given by,
\begin{equation}
p=p_{0} \left(\frac{\rho}{\rho_0}\right)^\gamma,
\label{eq:18}
\end{equation}
where, $p_0$ and $\rho_0$ are the pressure and density, respectively,
at the surface of the star and $\gamma$ is the polytropic index. Then 
writing the density and the Lorentz force term of Eq.~(\ref{eq:14}) as functions of 
radius and the radial velocity, we can easily obtain an expression for the gas velocity gradient as,
\begin{eqnarray}
\frac{dw}{dx}=\frac{w}{x}\frac{N(w,x)}{D(w,x)}~,
\label{eq:19}
\end{eqnarray}
where, $w=u/u_A$ is the gas speed normalised using the Alfv\'{e}n
speed and $x=r/r_A$, is the radial distance expressed in units of the
Alfv\'{e}n radius. The quantities $N(w,x)$ and $D(w,x)$ are the
numerator and denominator respectively and are given by,
\begin{eqnarray}
N(w,x)= \left(2 \gamma S_T (wx^2)^{1-\gamma} - \frac{S_G}{x}(1-\Gamma_d \cdot \Theta(x-x_d))\right) \nonumber\\
\times (wx^2-1)^3 +~ S_{\Omega}x^2(w-1)\left(1-3wx^2 + (wx^2+1)w \right)
\label{eq:20}
\end{eqnarray}
and
\begin{eqnarray}
D(w,x)= \left(w^2-\gamma S_T (wx^2)^{1-\gamma}\right)(wx^2-1)^3-
S_{\Omega}x^2 \times \nonumber\\
(wx^2)^2\left(\frac{1}{x^2}-1\right)^2.
\label{eq:21}
\end{eqnarray}
In the above equations, the parameters $S_T=\frac{2kT_A}{m_p u_A^2}$,
$S_G=\frac{GM_*}{r_A u_A^2}$ and $S_\Omega=\frac{\Omega^2
  r_A^2}{u_A^2}$ along with $\gamma$ uniquely determine the locations
of the critical points, and hence the morphology of the family of
solutions of Eq.~(\ref{eq:19}). The critical points are, as usual,
defined as the locations at which the both the numerator and
denominator vanish, thereby keeping the right-hand side of
Eq.~(\ref{eq:19}) finite. The presence of the Heaviside function in
Eq.~(\ref{eq:20}) represents the formation of dust at the location
$x=x_d$, the dust condensation radius in units of the Alfv\'{e}n
radius. The critical wind solution of Eq.~(\ref{eq:19}) will yield the
gas velocity profile and thereby enable determination of other
dependent variables, such as the dust velocity profile (to be
discussed below), the Mach number as a function of distance from the
star, the azimuthal velocity of the gas, the azimuthal component of
the magnetic field, the temperature profile and the density structure
of the gas in the envelope of the AGB star.

As mentioned above, we can determine the dust velocity profile after
having determined $u$, the gas velocity as a function of radius, with
the help of Eqs.~(\ref{eq:14}) and (\ref{eq:15}). This allows us to
express the drift speed as (e.g. \cite{Lamers}),
\begin{equation}
(v-u)^4+(v-u)^2 a_{th}^2 - \left( \frac{Q_{rp}L_*}{4 \pi r^2 \rho c} \right) ^2=0.
\label{eq:22}
\end{equation}
This yields the solution (after employing Eq.~(\ref{eq:19})),
\begin{equation}
v(r)=u(r)+\left(\frac{\sqrt{a_{th}^4+4\left(\frac{\Gamma_d GM_{*}}{\pi a^2 n_d r^2}\right)^2}-a_{th}^2}{2}\right)^{1/2}
\label{eq:23}
\end{equation}
The dust grain number density $n_d$, can be obtained from
Eq.~(\ref{eq:5}). In the current study we are not solving for the dust
velocity simultaneously with Eq.~(\ref{eq:19}). However, since the
dust-to-gas ratio is already a small number and decreases
monotonically in the wind away from the star to become even smaller,
we shall therefore make the simplifying assumption that $n_d m_d /
\rho \approx \langle\delta\rangle$, the average dust-to-gas ratio in the wind. This essentially
implies that $n_d$ falls off faster than $\rho$, in the wind; this is
a reasonable assumption since the dust velocity is expected to exceed
the gas velocity.

Finally the energy flux per second per steradian can be determined by
expressing Eq.~(\ref{eq:19}) as a total derivative. This yields,
\begin{eqnarray}
\frac{F}{\rho u r^2}&=& \left( \frac{u^2}{2} + \frac{u_{\phi}^2}{2} + \frac{\gamma}{\gamma - 1} \frac{p_A}{\rho_A} \left(\frac{\rho}{\rho_A}\right)^{\gamma-1} \right.\nonumber\\
& & ~~~ \left.- \frac{GM_{*} (1-\Gamma_d\Theta(r-r_d))}{r} 
 - \frac{B_{\phi}B_r}{4 \pi \rho} \frac{\Omega r}{u} \right)\!.
\label{eq:24}
\end{eqnarray}
It is immediately evident upon inspecting Eq.~(\ref{eq:24}) that the
Heaviside function will present a discontinuity in the flux at the
dust formation radius, therefore, in order to preserve the constancy
of energy flux across the dust formation interface at $r=r_d$, it
becomes necessary to subtract a constant term to the energy flux
outside the dust formation interface, such that,
\begin{equation}
\frac{\left.F(r) \right|_{{r=r_d}^{-}}}{\rho u r^2}=\frac{\left.F(r) \right|_{{r=r_d}^{+}}}{\rho u r^2} - \mathrm{constant}.
\label{eq:25}
\end{equation}
This constant is essentially the difference between the energy fluxes on either side of the dust formation radius ($r=r_d$) and is given by $const=GM_*\Gamma_d/r_d$. 
Such a constant term effectively redefines the gravitational
potential, without altering the dynamics; i.e., its derivative
vanishes, since it is a constant and thus, it does not change the
solution topology of Eq.~(\ref{eq:19}). Therefore we can write, 
\begin{eqnarray}
\frac{F(r \leq r_d)}{\rho u r^2} &=& \left( \frac{u^2}{2} +
  \frac{u_{\phi}^2}{2} +
  \frac{\gamma}{\gamma - 1} \frac{p_A}{\rho_A}
  \left(\frac{\rho}{\rho_A}\right)^{\gamma-1} \right.\nonumber\\
& & ~~~
\left.- \frac{GM_{*}}{r} 
 - \frac{B_{\phi}B_r}{4 \pi \rho} \frac{\Omega r}{u} + \frac{GM_*\Gamma_d}{r_d}\right).
\label{eq:26}
\end{eqnarray}
Similarly, for $r \geq r_d$ we obtain Eq.~(\ref{eq:24}), thus we
ensure that flux is constant across the dust formation interface.

This completes our derivation of the governing equations for the
hybrid wind model. We see that once a solution of Eq.~(\ref{eq:19}) is
determined, i.e., the gas velocity profile, it in turn determines all
the other relevant variables including the dust grain velocity as
given by Eq.~(\ref{eq:23}) and the energy fluxes given by
Eqs.~(\ref{eq:24}-\ref{eq:26}). In the following section we shall
describe the numerical treatment breifly and the results are presented
and discussed in \S\ref{sec:Results and Discussion}.

\section{Numerical details}
\label{sec:numer}
The ordinary differential equation (ODE) for the gas velocity
gradient, Eq.~(\ref{eq:19}) was integrated as an initial value problem
for a range of initial conditions in the $w-x$ phase space. The domain
of integration was $x_0 \leq x \leq 5$, where $x_0$ represents the
stellar surface. Table~\ref{tab:parameters} summarises all the
physical parameters employed for a typical AGB star. The first step in
the numerical procedure is the determination of the critical points,
this is described below.
\begin{table*}
 \centering
\label{tab:parameters}
 \begin{minipage}{140mm}
  \caption{Summary of the different parameters for modelling an AGB star hybrid wind}
  \begin{tabular}{@{}lcl@{}}
  \hline
   Parameter     & Symbol        &  Value and/or Comment \\
 \hline
Mass & $M$ & $\sim 5M_{\odot}$ \\
Radius & $R_0$ & $\sim 500R_{\odot}$ \\
Mass loss rate & $\dot{M}$ & $\sim 1.6 \times 10^{-6} M_{\odot}$ per year \\
Surface magnetic field strength & $B_0$ & $\sim 1$ G \\
Bulk gas velocity (radial) at the surface & $u_0$ & $\sim 2 \times 10^{-8} v_{esc,0}$ (vanishingly small) \\
Surface temperature (effective) & $T_0$ & $\sim 3000$K \\
Stellar rotation rate & $\Omega$ & $\sim 2 \times 10^{-10}$ rad/s \\
Surface escape velocity & $v_{esc,0}$ & $6.19 \times 10^6$ cm/s \\
Polytropic exponent & $\gamma$ & 1.06 \\
\hline
\end{tabular}
\end{minipage}
\end{table*}
It is to be mentioned in this regard, that we chose the value of the polytropic exponent to be approximately mid-way between unity and the values employes in current 2D-axisymmetric MHD codes, $\approx 1.13$  \citep[e.g.][]{Keppens1999}. A value of unity represents an insothermal equation of state and since the envelopes of AGB stars may be well-mixed due to convection, we chose a value slightly more than unity to resemble a sort of effective cooling.

\subsection{Determination of critical points}\label{sec:critpts}
The critical points are the locations in the $w-x$ phase space at
which both the numerator ($N(w,x)$) and the denominator ($D(w,x)$) in
Eqs.~(\ref{eq:20}) and (\ref{eq:21}), identically vanish. Once the
values of the parameters $S_T$, $S_G$, $S_\Omega$ and $\gamma$ are
established, we then proceed to solve the system of non-linear
algebraic equations given by,
\begin{eqnarray}
N(w_s,x_s)=0 \label{eq:27}\\
N(w_f,x_f)=0\label{eq:28}\\
D(w_s,x_s)=0\label{eq:29}\\
D(w_f,x_f)=0,
\label{eq:30}
\end{eqnarray}
where, $x_s$ represents the distance from the photosphere (in units of
$r_A$) at which the gas velocity is equal to the local sound speed,
$w_s$ (in units of $u_A$). Similarly, the point $(w_f,x_f)$ represents
the location in the phase space at which the kinetic energy density of
the gas is equal to the local total magnetic energy density, i.e.,
$\frac{1}{2}\rho u^2 = \frac{B_r^2+B_{\phi}^2}{8\pi}$; the so-called
fast point. The root finding is accomplished using a
Levenberg-Marquardt medium-scale root finding algorithm
\citep[e.g.][]{More1977}. Typical tolerances employed were about
$10^{-15}$ in order to ascertain the zeros of the system of equations
(\ref{eq:27} - \ref{eq:30}). This procedure is carried out for
parameters $S_T$, $S_G$ and $S_\Omega$ on both sides of the dust
formation interface. Across the interface the only change is that,
\begin{equation}
S_G^+=S_G^-(1-\Gamma_d),
\label{eq:31}
\end{equation}
where $S_G^+$ represents the value of the parameter outside of the
dust formation interface, while $S_G^-$ represents the value inside
the dust formation interface. The remaining two parameters $S_T$ and
$S_\Omega$ are unchanged across the interface. Presently, we describe
the procedure employed for determining the location of the radial
Alfv\'{e}n point and the dust parameter $\Gamma_d$.

\subsection{Determination of the radial Alfv\'{e}n point and dust parameter $\Gamma_d$}\label{sec:ua_ra}

We begin with a set of parameters $S_T$, $S_G$ and $S_\Omega$ that are
chosen arbitrarily; however, with the constraint that, for the given
set of parameters, a critical solution is not physically possible. The
rationale being that a purely Weber-Davis wind is not possible for an
AGB star. This will be explained in detail later, when the results are
discussed. Once the above mentioned parameters are chosen, the
remaining parameters $u_A$ and $r_A$ are continuously varied for
different values of $\Gamma_d$ until we are able to achieve a physical
critical solution. The chief criterion for the latter being that the
solution is continuous through the radial Alfv\'{e}n point and does
not have a kink and is required to originate at the base of the wind
sub-sonic, pass through all three critical points, viz., the sonic
point, the radial Alfv\'{e}n point and the fast point and subsequently
leave the star super-Alfv\'{e}nic. The reader at this stage is
referred to the subsection on determination of the critical solution
for details (see \S\ref{sec:crit_sol}). Following an initial guess
for the parameters ($u_A$, $r_A$) with $\Gamma_d$ fixed to a certain
value, they are varied with typical step sizes of $10^{-6}$ until
suitable values are obtained. Once $u_A$ and $r_A$ are determined, the
temperature at the radial Alfv\'{e}n point is determined according to,
\begin{equation}
T_A=T_0\left(\frac{u_0 R_0^2}{u_Ar_A^2}\right)^{1-\gamma},
\label{eq:32}
\end{equation}
wherein, the parameters at the base of the wind (subscripted with $0$)
are given in Table~\ref{tab:parameters}. We now turn our attention to the matter of
integrating the ODE in Eq.~(\ref{eq:19}) after having determined the
above mentioned parameters.

\subsection{Integration of the ODE and determination of the critical solution}\label{sec:odeint}

\subsubsection{ODE Integration}\label{sec:integration}

Integration of the ODE was accomplished with the software package
ODEPACK employing the subroutine DLSODE using backward difference
formulae and chord iteration with the Jacobian supplied
\citep{Hindmarsh1983,Hindmarsh1989}. Initial conditions were supplied
at the beginning of the integration. Typical error tolerances for
convergence testing that were employed were on the order of $10^{-12}$
for both the absolute and relative errors
\citep[see][]{Hindmarsh1983}. For a typical integration over the
domain $x_0 \leq x \leq 5$, a step size of $10^{-8}$ was employed,
resulting in typically $10^8$-$10^9$ function evaluations. A hybrid
stellar wind software package was specifically developed for the
current study and this was constructed to be capable of reproducing
self-consistently the entire family of solutions beginning with an
arbitrary choice of wind parameters. The entire code takes
approximately an hour to execute on an AMD Opteron$^{\textregistered}$
844~1.8~GHz~processor.

\subsubsection{\label{sec:crit_sol}The critical solution}

The critical solution is rather unique; it passes through all three
critical points. In the current study the critical solution was
determined using a tail procedure. For a given set of parameters
$\gamma$, $S_T$, $S_G$ and $S_\Omega$, forward and backward
integrations were carried out from the points $(w_s,x_s)$ and
$(w_f,x_f)$. The backward integration from $(w_s,x_s)$ was carried out
all the way to the photosphere of the star and similarly, the forward
integration from $(w_f,x_f)$ was carried out all the way to the outer
boundary of the domain at $x=5$. In between the backward integration
from $(w_f,x_f)$ was matched with the forward integration from
$(w_s,x_s)$. An initial guess was made for the matching point to be
half way between $x_s$ and $x_f$ defined by $x_m$. The tails of the
forward and backward integrations were terminated at this location and
the values of the gas velocity $w$ and the velocity gradient
$\frac{dw}{dx}$ were compared for the two tails to ensure that the
conditions
\begin{eqnarray}
\left|
\left(
\Delta w=\left.w(x_m) \right|_{{x=x_m}^{-}} - \left.w(x_m) \right|_{{x=x_m}^{+}}\right)\right|
&\leq 10^{-7}\label{eq:33}\\
\left|
\left(
\left.\frac{dw}{dx} \right|_{{x=x_m}^{-}} - \left.\frac{dw}{dx} \right|_{{x=x_m}^{+}}\right)\right|
&\leq 10^{-7},
\label{eq:34}
\end{eqnarray}
were both met. If the conditions in Eqs.(\ref{eq:33}) and
(\ref{eq:34}) were not met, then depending upon the sign of $\Delta
w$, the matching point $x_m$ was appropriately shifted either forward
or backward by a small amount, typically by $\Delta x_m=10^{-6}$ and
the tail procedure was re-iterated. Once the tail procedure was
successful the critical solution was considered to be determined. This
ensured that the critical solution was continuous through the radial
Alfv\'{e}n point.  The tolerance employed in Eqs.~(\ref{eq:33}) and
(\ref{eq:34}) was considered to be sufficient given the fact that the
integration step size was $\Delta x = 10^{-8}$. For achieving a higher
tolerance, a further reduction in the step size was found to be
necessary, rendering the the procedure needlessly lengthy and
increasingly cumbersome in a computational sense.

This completes our discussion of the numerical details regarding
determining the complete family of solutions of Eq.~(\ref{eq:19}). The
results are presented in the following section and are discussed
therein.

\section{Results and Discussion}
\label{sec:Results and Discussion}
We present the results obtained by integrating the ODE in
Eq.~(\ref{eq:19}) to obtain the family of solutions. In the current
study, the following methodology was employed for calculating the
hybrid wind. First, for a certain set of arbitrary parameters \{$S_T$,
$S_G$, $S_\Omega$ and $\gamma$\}, Eqs.~(\ref{eq:27}-\ref{eq:30}) were
solved to obtain the location of the slow and fast points, with the
radial Alfv\'en point located at $(1,1)$ in the $w-x$ phase space. At
this stage, the dust parameter $\Gamma_d$ was set equal to zero,
meaning that there hasn't been any dust formation in the gas. The parameters are chosen such that a pure WD wind is not a physical one, i.e., it is not continuous through the radial Alfv\'en point (see Figure~\ref{fig:figure3} and discussion thereof). The rationale being that for AGB stars, it is not possible to have a mass efflux without dust formation in the envelope, therefore a pure WD wind is explicitly required to not be a physical solution. Second,
the dust parameter was set equal to a fraction such that $0 \leq
\Gamma_d \leq 1$ and the procedure described in \S\ref{sec:ua_ra} for
determing the Alfv\'en velocity and radius is carried out in tandem
with solving Eqs.~(\ref{eq:27}-\ref{eq:30}) to obtain the sonic point
and the fast point. Each time this yields a set of parameters \{$u_A$,
$r_A$, $(w_s,x_s)$, $(w_f,x_f)$\}. With these parameters,
Eq.~(\ref{eq:19}) is integrated to obtain the critical solution
according to the procedure described in \S\ref{sec:crit_sol}. If
success is achieved in finding the critical solution then the
iterations are ceased and the resulting parameters are fixed for the
given value of $\Gamma_d$.

We then determine the temperature profile using the velocity profile
of the critical solution. This is achieved using the prescription,
\begin{equation}
T(r)=T_A\left(\frac{u_A r_A^2}{ur^2}\right)^{\gamma-1},
\label{eq:35}
\end{equation}
where, $T_A$ is the Alfv\'en temperature and is given by
Eq.~(\ref{eq:32}). Once the temperature profile is known, it is
possible to invert Eq.~(\ref{eq:35}) to yield the radius $r_d$, at
which the temperature falls below the dust condensation temperature
$T_d$. Once the dust condensation radius ($r_d$) is determined, we then
proceed to determine the family of solutions to Eq.~(\ref{eq:19}) such
that integration from the photoshpere at $r=R_0$ to $r=r_d$ is carried
out with $\Gamma_d=0$ and integration from $r=r_d$ to the outer
boundary at $r=5r_A$ is done with $\Gamma_d \neq 0$. The same is then
done for the critical solution as well.

\begin{figure}
\begin{center}
\includegraphics[width=3in, scale=0.8]{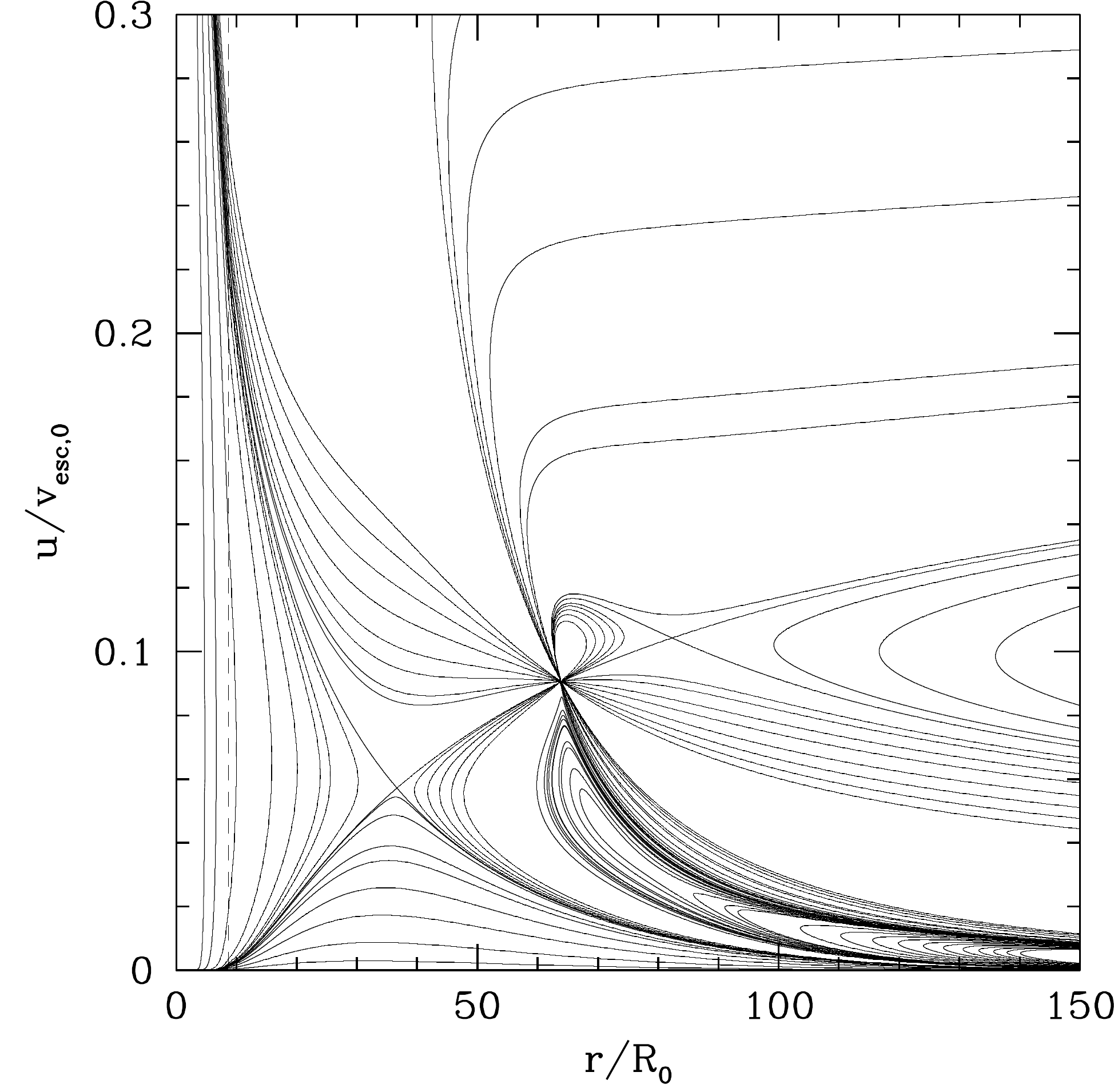}
\end{center}
\caption{Family of solutions of Eq.~(\ref{eq:19}) with parameters $u_A \approx 0.09v_{esc,0}$, $r_A \approx 63.93R_0$, $\Gamma_d=0.3$ and remaining parameters as given in Table 1. The dashed line at $r \approx 8.65 R_0$ represents the dust formation radius.} 
\label{fig:figure1}
\end{figure}

As mentioned before, it is implicitly assumed, for the sake of
analysing a patently simple model, that beyond the dust condensation
radius the value of $\Gamma_d$ is constant. Concordantly, all the dust
forms at the condensation radius and the dust-to-gas ratio is small and given by
Eq.~(\ref{eq:5}). The family of solutions determined using the
procedures described above are shown in Figure~\ref{fig:figure1}. For
the solutions shown therein, the dust parameter was fixed at
$\Gamma_d=0.3$.
\begin{figure}
\begin{center}
\includegraphics[width=3in, scale=0.8]{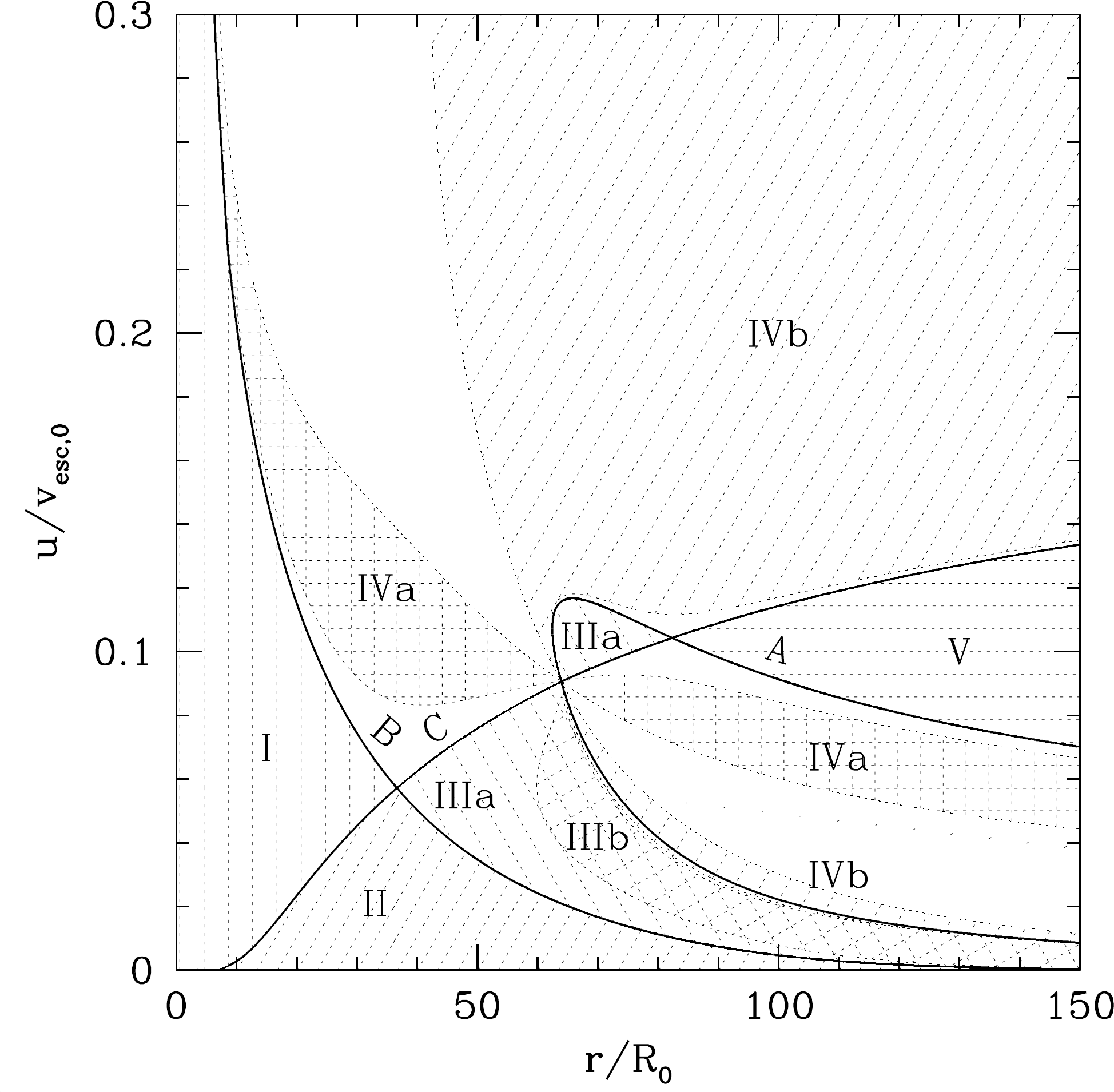}
\end{center}
\caption{Illustration showing the locations of the different types of solutions of Eq.~(\ref{eq:19})} 
\label{fig:figure2}
\end{figure}
It can be seen that the dust condensation radius is located at
$r_d\approx 8.65 R_0$. The temperature at this location was determined
using Eq.~(\ref{eq:35}), to be approximately $1200$K. The dust
condensation temperature was chosen somewhat arbitrarily to lie
between $1000 - 1500$K. Within the dust condensation radius $x_d$, the
wind solution to Eq.~(\ref{eq:19}) is a purely Weber-Davis-type of
wind, and thereafter it is a hybrid wind with dust grains
included. The topology of solutions in Figure 1 looks typically like a
WD-solution, with the three critical points clearly visible, viz., the
sonic point, the radial Alfv\'en point and the fast point. The
critical solution emerges from the surface of the star sub-sonic, gets
accelerated through to the dust condensation radius, subsequently
passes through the three critical points and finally emerges
super-Alfv\'enic at large distances.

At this stage it is convenient to classify the different types of
solutions in Figure~\ref{fig:figure1}. The different types of
solutions and their respective locations in the phase space are
illustrated qualitatively in Figure~\ref{fig:figure2}. Therein, it can be seen that the unphysical
double-valued solutions to the left of the sonic point are referred to,
in the current study, as Type I solutions. The failed wind solutions,
directly below the sonic point, are called Type II. The unphysical
multi-valued solutions that make loops, between the radial Alfv\'en
point and the fast point, are designated as Type IIIa. The
unphysical double-valued solutions adjacent to the loop solutions are
called Type IIIb.

The solitary unphysical solution that passes through the sonic X-type
singularity, is called the \emph{Bondi solution}, this is labelled as $B$,
in Figure~\ref{fig:figure2}. While the unphysical double-valued solution that
intersects the critical solution, at exactly the two Alfv\'en points, is
designated the \emph{Alfv\'enic solution}; labelled $A$. Similarly,
the critical solution is labelled as $C$.

The unphysical wind solutions that start at the photophere with
super-sonic velocities, just to the right of the Bondi solution and
subsequently pass through the radial Alfv\'en point and get
decelerated to sub-sonic velocities at large distances from the star,
are designated as Type IVa. Meanwhile, the unphysical double-valued
solutions that pass through the radial Alfv\'en point alone, are
designated as Type IVb. Finally, the unphysical double-valued
solutions, in the region between the critical solution and the
Alfv\'enic solution, immediately to the right of the fast point, are
referred to as Type V.

As mentioned earlier, the dust grains condense from the gas beyond
$r_d$, where the temperature falls below the condensation
temperature. Beyond $r_d$, we threfore solve the hybrid ODE with a
fixed value of $\Gamma_d$. Thus at the dust condensation radius, the
two types of solutions, dust-free and hybrid, must match, in terms of
velocity of the gas. This is illustrated in Figure~\ref{fig:figure3},
where we have expressed the coordinates using a logarithmic scale to
facilitate examination.
\begin{figure}
\begin{center}
\includegraphics[width=3in, scale=1.0]{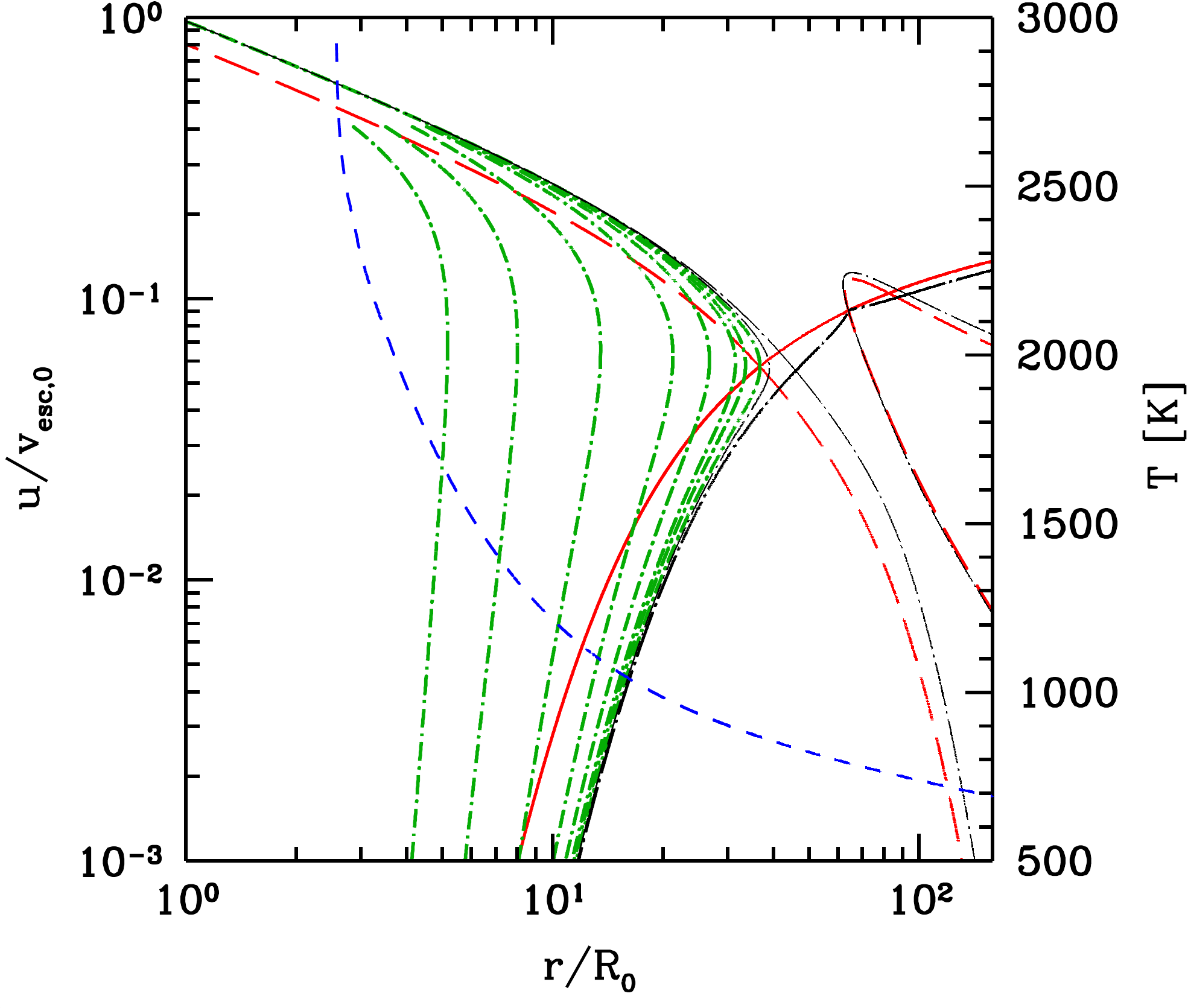}
\end{center}
\caption{Plausible hybrid wind solutions with parameters $u_A \approx
  0.09v_{esc,0}$, $r_A \approx 63.93R_0$, $\Gamma_d=0.3$ and remaining
  parameters as given in Table 1. The red solid line and black
  long-dash-dotted line intersecting at the radial Alfv\'en point, are
  the critical solutions of the hybrid wind model and pure WD wind
  model, respectively. The green-short-dash-dotted lines are possible
  Type I wind solutions of a pure WD wind that can leave the star via
  the hybrid critical solution after dust condensation at the
  intersection points of the green-short-dash-dotted lines and the red
  solid line. The blue dashed line is the temperature profile
  calculated for the hybrid wind critical solution. It should be
  interpreted using the secondary axis.}
\label{fig:figure3}
\end{figure}
In Figure~\ref{fig:figure3}, the red solid line represents a hybrid
wind solution for a dust parameter of $\Gamma_d=0.3$, while the black
long-dash-dotted line that passes through the radial Alfv\'en point
with a kink at $(u_A,r_A)$, is a pure WD solution with
$\Gamma_d=0$. As can be clearly seen, the kink at the radial Alfv\'en
point indicates that this solution is not physical. This indicates that it is not possible to have a pure WD wind for AGB stars; only with dust formation is it possible to achieve an outflow. The Bondi type
solutions for both the hybrid wind (long-dashed red line) and the pure
WD solution (black long-dash-dotted line), can also be seen to pass
through the respective sonic points. Similarly the two fast points can
also be distinguished clearly for the two types of solutions; the
Alfv\'enic solutions pass through them. It is to be mentioned at this
juncture, that inclusion of an outward force in the wind due to
radiation pressure has the effect of suppressing both the sonic point
and the fast point towards the photosphere; the sonic and the fast
points for the hybrid wind (in red) can clearly be seen to lie inside
their respective counterparts of the pure WD wind, in terms of
distance from the star's photosphere. The green short-dash-dotted
lines are solutions of Type I, for a pure WD wind. The points of
intersection of the red solid line representing the critical solution
of the hybrid model and the green short-dash-dotted lines, represent
different dust formation radii. The temperatures corresponding to
these can be inferred using the blue line, which represents the temperature
profile, determined using the critical solution of the hybrid wind, via
Eq.~(\ref{eq:35}). Thus, if a wind solution starts off at the base of
the wind sub-sonic and travels through the envelope of the AGB star
according to (say) the fourth green short-dash-dotted line from the
left, then it can pass through the dust formation radius at
approximately $1000$K, just ahead of about $10 R_0$ and leave the star
via the red solid line, the critical solution of the hybrid wind. We
therefore see that the unphysical Type I solutions that start off
dust-free, can intersect the hybrid critical solution and leave the
star as a hybrid critical wind. However, there is a constraint. The
last Type I solution that can leave the star as a hybrid wind
intersects the critical solution in red, at the sonic point. Any
solution of Type I that intersects the red solid line, after it has
turned and become double-valued, does not represent a physical hybrid
wind. This is illustrated in the black long-dash-dotted Type I
solution that turns and then intersects the red solid line, just ahead
of the hybrid sonic point in Figure 3. Therefore, all the green
short-dash-dotted lines are allowed possible solutions of different
hybrid winds.

Figure~\ref{fig:figure3} also shows that if a hybrid wind is to be
achieved, there are two possibile scenarios. First, the dust formation radius can lie within or
at most upon the sonic point of the hybrid wind solution. Second, it can lie beyond or upon the fast
point of the hybrid solution, or exactly at the radial Alfv\'en point. Only the first
possibility places the dust formation temperature in the acceptable
range for typical AGB parameters so we will focus mainly on this class
of solutions (however, c.f. Figure~\ref{fig:figure9} and discussion thereof).
While the latter scenarios do represent legitimate mathematical solutions, it is however unlikely
that the dust formation temperature be significantly lower than about
$1000$K and concomitantly, that the dust formation radius in AGB
stars, should lie as far out as nearly $70 R_0$. Since it is expected
that the dust formation must occur within a few stellar radii in AGB
stars, the only plausible solutions are therefore the Type I green
short-dash-dotted lines, that intersect the hybrid critical line in
red.

\begin{figure}
\begin{center}
\includegraphics[width=3in, scale=0.8]{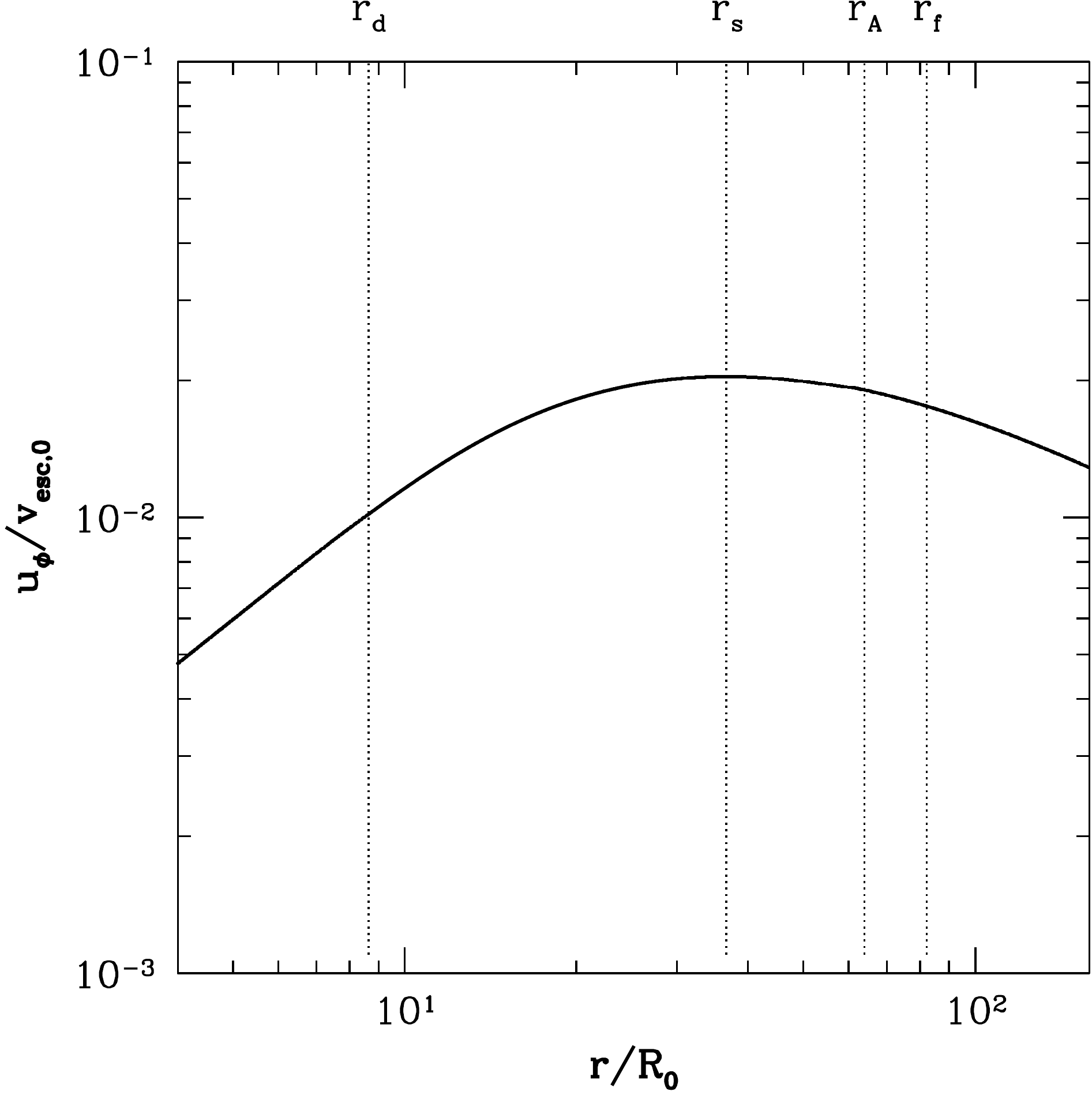}
\end{center}
\caption{Critical azimuthal gas velocity as a function of radius. The
  vertical dotted lines indicate the locations of the dust
  formation radius and the the three critical points.}
\label{fig:figure4}
\end{figure}

For the hybrid critical solution shown in Figures~\ref{fig:figure1}
and~\ref{fig:figure3}, it is possible to calculate the azimuthal
velocity profile of the gas by manipulating Eq.~(\ref{eq:17}). This is
carried out and is shown in Figure~\ref{fig:figure4}. As can be seen the gas velocity
profile rises sharply to a maximum value close to the sonic point and
then falls off less steeply than the rise; this is consistent with the usual
WD-picture.
The locations of the dust formation radius and the three critical
points are also indicated therein.

\begin{figure}
\begin{center}
\includegraphics[width=3in, scale=0.8]{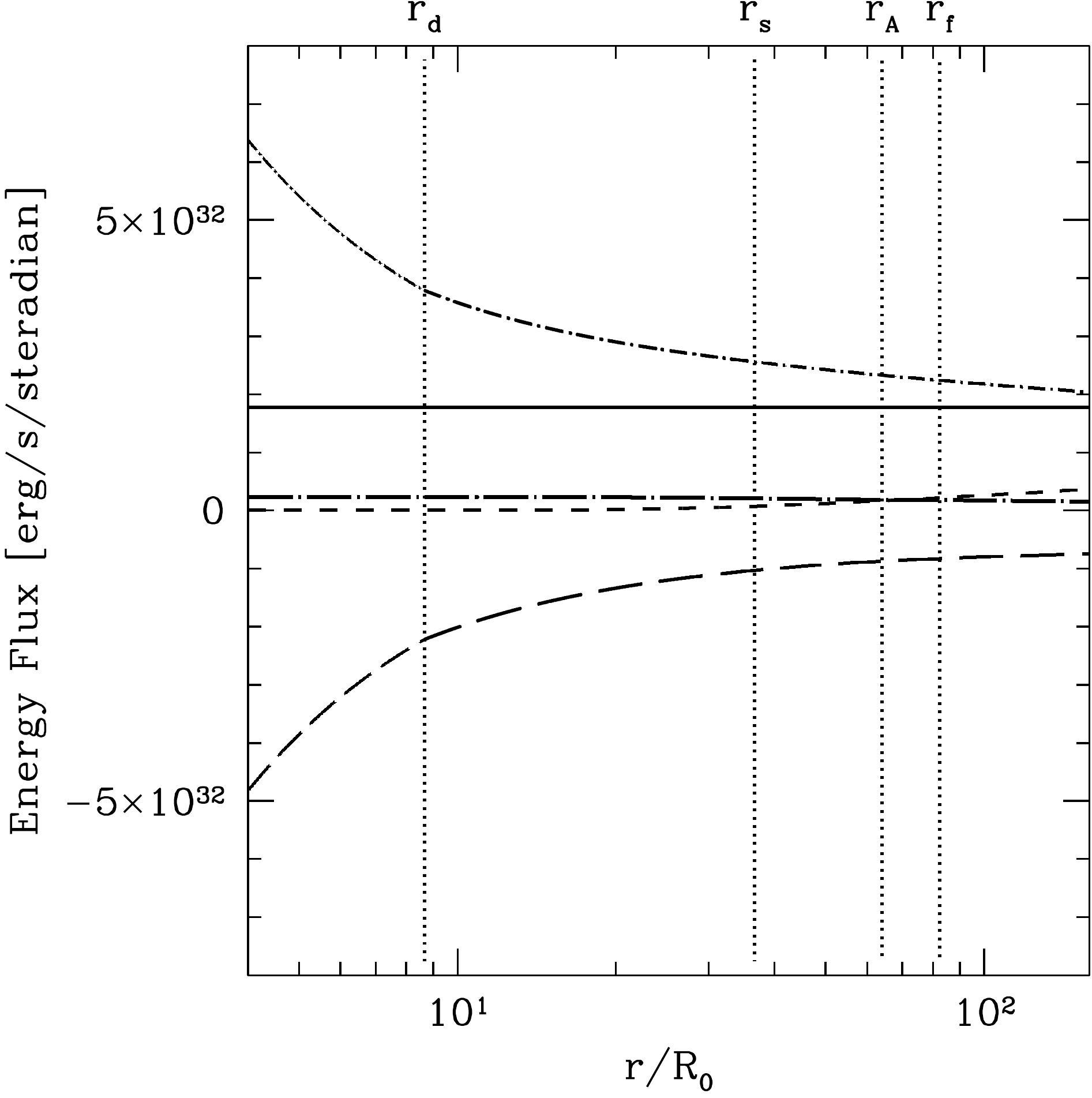}
\end{center}
\caption{Plot showing the energy fluxes calculated using
  Eqs.~(\ref{eq:24}-\ref{eq:26}) for the critical hybrid wind solution
  with parameters $u_A \approx 0.09v_{esc,0}$, $r_A \approx 63.93R_0$,
  $\Gamma_d=0.3$ and remaining parameters as given in Table 1. The
  vertical dotted lines indicate the dust formation radius and the the
  three critical points. The solid horizontal line represents the
  total constant energy flux.  From the top short-dash-dotted curve
  shows the magneto-rotational energy, the long-dash-dotted curve
  represents the enthalpy, the short-dashed curve shows the variation
  in kinetic energy and finally at the bottom the long-dashed line
  represents the gravitational potential energy of the gas.}
\label{fig:figure5}
\end{figure}

In Figure~\ref{fig:figure5} we have plotted the energy flux per second
per steradian leaving the star, in the equatorial plane, for the
critical solution, alongside the various components of the energy
flux.
The solid line shows the total energy flux in the gas as function of
radius; it is constant for the critical solution.  It
is the sum of all the other lines on the plot. This is calculated via
Eqs.~(\ref{eq:24}-\ref{eq:26}). The short-dash-dotted line is the
magneto-rotational energy flux, which is the sum of the last two terms
in Eq.~(\ref{eq:24}). As the distance from the photosphere increases,
the radial magnetic field falls of as $\sim 1/r^2$, however the
rotational energy is expected to peak close to the sonic point, as can
be inferred from Figure~\ref{fig:figure4}, where the azimuthal
velocity peaks. The two competing terms collectively yield a large
positive sum closer to the star, and the contribution diminishes quite
rapidly, beyond approximately the dust formation radius, as can be seen
in Figure 5. The long-dashed line represents the gravitational
potential energy of the wind, the third term in Eq.~(\ref{eq:24}) and
it is therefore negative. The short-dashed line is the kinetic energy
of the wind (the first term in Eq.~(\ref{eq:24})) which gradually
increases away from the star as the wind is accelerated. On the other
hand, the long-dash-dotted line is the enthalpy (the second term in
Eq.~(\ref{eq:24})) which gradually decreases from a maximum value at
the stellar surface where the density of the gas is expected to be the
highest and follows a power law, of the form $\sim
\rho^{\gamma-1}$. The decrease is quite gradual as the exponent is
$0.06$ since we used $\gamma=1.06$. The contribution of the kinetic
energy and the enthalpy are quite small in comparison to the
magneto-rotational energy. However, when the magneto-rotational energy
is added to the gravitational potential energy, then the sum is
comparable to the kinetic energy. Again, the dust formation radius and
the three critical points are indicated in the figure. It is also to be mentioned 
that, since it is possible to recover the momentum terms of Eq.~(\ref{eq:14})
by differentiating Eq.~(\ref{eq:24}), therefore the slopes of the different 
lines shown in Figure 5 would indicate the contributions of these terms. 
In particular, it can be clearly seen that the slopes of the magneto-rotational
energy and the gravitational energy (which includes radiation pressure) 
are the most prominent features in Figure 5. This indicates that the the most
significant contributions to accelerating the wind are indeed due to the Lorentz force term
plus rotational term in Eq.~(\ref{eq:14}) and the gravitational potential which includes
the effect of radiation pressure on the dust grains. In this regard, it is evident upon inspecting
Figure 5, that beyond approximately the radial Alfv\'en point, the gravitational energy flux becomes
flat, indicating that the acceleration due to the gravitational potential modified by radiation
pressure, becomes negligible in the wind. However, beyond $r=r_A$, the magneto-rotational energy still has a small contribution and continues to accelerate the wind. In summary, at small distances, ($r  < \approx r_s$), both the magneto-rotational terms and the modified gravitational potential terms of Eq.~(\ref{eq:14}) have a combined and pronounced effect in accelerating the wind; however, at large distances ($r > \approx r_A$) the effect of the gravitational potential becomes negligible while there still persists a small contribution from the magneto-rotational term, that mildly accelerates the wind outward.

\begin{figure}
\begin{center}
\includegraphics[width=3in, scale=0.8]{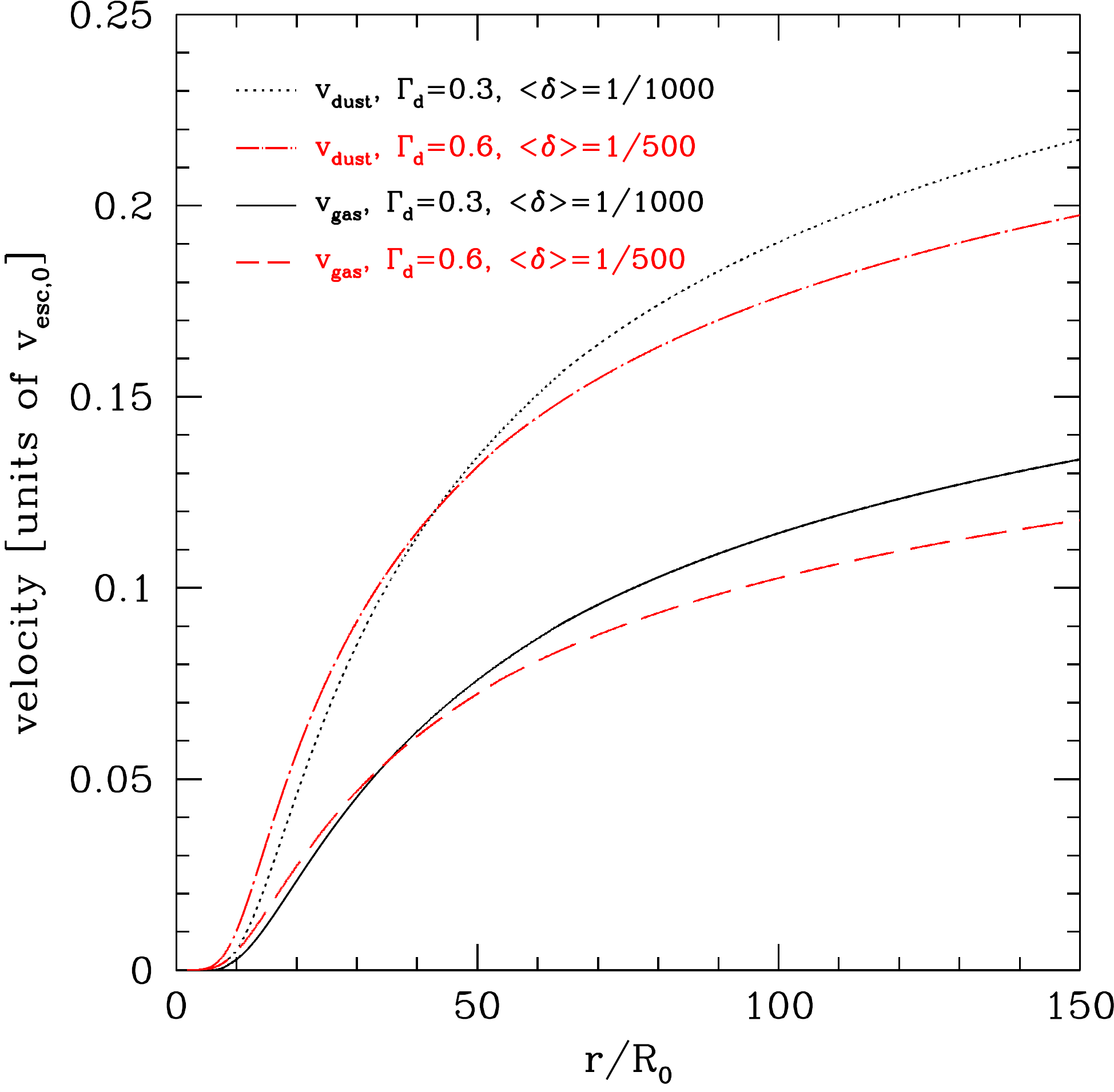}
\end{center}
\caption{Dust and gas velocity profiles for two hybrid winds with
  parameters $u_A \approx 0.09v_{esc,0}$, $r_A \approx 63.93R_0$,
  $\Gamma_d=0.3$ (in black) and with parameters $u_A \approx 0.07v_{esc,0}$, $r_A \approx 53.05R_0$, $\Gamma_d=0.6$ (in red). In the former case, the dust-to-gas ratio was $\langle \delta \rangle = 1/1000$ and in the latter case it was double this value, i.e. $\langle \delta \rangle = 1/500$. The remaining wind parameters are given in Table 1. The dust velocity profile in each case is determined using Eq.~(\ref{eq:23}).}
\label{fig:figure6}
\end{figure}

Figure~\ref{fig:figure6} shows the velocity profiles of the dust and
the gas plotted as a function of distance from the centre of the star.
The velocity of the dust grains exceeds that of the gas bulk, as is
required by the hybrid wind model, in order to produce an outward drag
force on the gas, to accelerate it outward. The velocity of the dust
grains is determined according to Eq.~(\ref{eq:23}), once the velocity
profile of the gas is ascertained, with an assumed average value for
the dust-to-gas ratio $\langle\delta\rangle$. Figure 6 shows the gas and 
dust velocity profiles for two different sets of model parameters. The solid
black line represents the gas velocity while the dotted black line represents 
the corresponding dust velocity. This model had parameters $\Gamma_d=0.3$
and $\langle \delta \rangle = 1/1000$. In order to investigate the effect of changing
the average dust-to-gas ratio, we kept all other parameters of the model fixed,
in particular, the radiation pressure mean efficiency and the stellar luminosity,
were kept constant. Then, according to Eq.~(\ref{eq:15}), if the average dust-to-gas 
ratio is doubled then, then accordingly, the dust parameter $\Gamma_d$ must also double. Thus, for the second model's results shown in Figure 6, we took $\Gamma_d=0.6$ and
$\langle \delta \rangle = 1/500$. Thus, the red-dashed line represents the gas velocity profile, 
while the red-long-dash-dotted line represents the corresponding dust velocity profile.
However, changing the dust parameter also changes the locations of the the three critical points.
With $\Gamma_d=0.3$ the Alfv\'en velocity and Alfv\'en radius were found to be $u_A \approx 0.09v_{esc,0}$, $r_A \approx 63.93R_0$, while for $\Gamma_d=0.6$, the corresponding values were found to be $u_A \approx 0.08v_{esc,0}$, $r_A \approx 53.05R_0$ respectively. It can clearly be seen that within about $50R_0$ (which is approximately the location of the radial Alfv\'en point for the second model), the dust in the second model (red-long-dash-dotted line) has a steeper rise, indicating a larger acceleration in the wind. Beyond about $50R_0$, the acceleration of the wind in the second model ($\Gamma_d=0.6$) starts to decline (see Figure 5 and discussion thereof). However, the wind in the first model ($\Gamma_d=0.3$) at this distance, is still getting accelerated, therefore it's velocity increases. Thus, when $\Gamma_d$ is smaller, acceleration due to radiation pressure continues to have an effect, out to larger distances from the star. In addition, at a distance of about $50R_0$, the temperature in the wind is about $\approx 800K$ for the model with $\Gamma_d=0.3$ and about 
$\approx 550K$ for the model with $\Gamma_d=0.6$. Thus, the bulk of the gas is much cooler in the latter case. As a result, it is natural that the velocities in the second model are slightly lower than the first. That being said, it is to be acknowledged that by increasing the dust parameter by increasing the average dust-to-gas ratio, the wind gets accelerated much faster closer to the star. This is expected; however, the terminal velocity in this case is lower, as the acceleration due to radiation pressure does not have a pronounced effect out to large distances ($r > \approx r_A$). 

\begin{figure}
\begin{center}
\includegraphics[width=3in, scale=0.8]{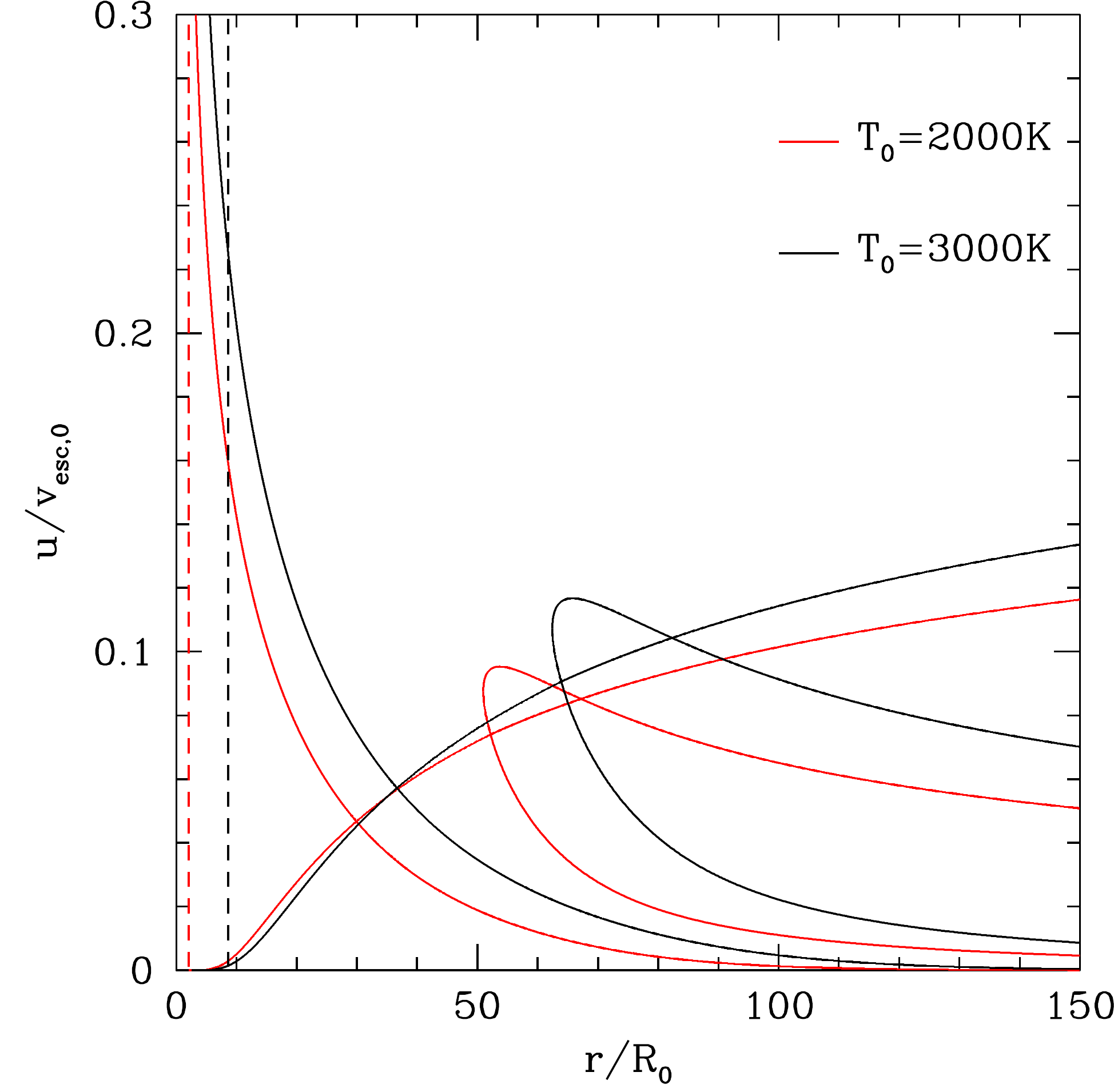}
\end{center}
\caption{Plot showing the effect of changing the temperature at the
  base of the stellar wind. The black lines are for a hybrid wind with
  parameters $T_0=3000$K, $u_A \approx 0.09v_{esc,0}$, $r_A \approx
  63.93R_0$ and $\Gamma_d=0.3$ while the red lines are for a hybrid
  wind with parameters $T_0=2000$K, $u_A \approx 0.07v_{esc,0}$, $r_A
  \approx 52.19R_0$ and $\Gamma_d \approx 0.62$. Both winds have
  identical remaining wind parameters, as given in Table 1. The dashed
  lines represent the locations of the respective dust formation radii
  for the two hybrid winds with $T_d \approx 1200$K, in both
  cases. Dust formation occurs closer to the stellar surface when the
  base temperature is lowered.}
\label{fig:figure7}
\end{figure}

At this stage we turn our attention to the question of changing the
temperature at the base of the wind. In AGB stars, it is likely that
due to density pulsations within the star, the temperature and density
at the stellar photosphere are likely to undergo change
\citep[e.g.][]{Lamers}. It has also been suggested by
\citet{Soker1999} that magnetic-cool spots probably exist in the
region around the equator of an AGB star, much like our sun. If such
is the case, then the temperature at the base of the hybrid wind will
change locally, and this will have an effect on the stellar wind beyond
the photosphere in the AGB envelope. In order to investigate this,
we constructed a hybrid model with an altogether different base
temperature. Following \citet{Soker1999}, we set the base temperature
in a magnetic-cool spot to be significantly less than the prescribed
average $3000$K, of the stellar photosphere. Accordingly, in order to
achieve a hybrid stellar wind, that starts with negligible velocity at
the stellar surface and gets accelerated to super-Alfv\'enic
velocities having passed through the dust formation radius and the the
three critical points, we treated $\Gamma_d$, $u_A$ and $r_A$ as free
parameters. As described in \S\ref{sec:ua_ra} we varied the values of
these parameters until a successful critical solution was
achieved. When the base temperature is changed to $2000$K, we found
that in order to achieve a hybrid stellar wind, the three free parameters
required to take on values; $\Gamma_d \approx 0.62$, $u_A \approx
0.07v_{esc,0}$ and $r_A \approx 52.19R_0$. Figure~\ref{fig:figure7}
shows how the morphology of the solution changes when the base
temperature is changed. We have merely shown the key solutions,
namely, the critical solutions, the Bondi and the Alfv\'enic
solutions.
The solid black lines are the hybrid wind solutions when the base
temperature is $T_0=3000$K and the dust parameter is
$\Gamma_d=0.3$. The lighter red lines are for $T_0=2000$K and
$\Gamma_d \approx 0.62$. The dashed lines at the far left indicate the
locations of the respective dust formation radii for the two models;
in both cases the dust formation temperature was taken to be $T_d
\approx 1200$K. It can be clearly seen in Figure 7, that when the base
temperature is decreased to $2000$K, all the critical points as well
as the dust formation radius, are suppressed towards the
photosphere. The formation of dust closer to the stellar surface is
directly related to the steep drop in the temperature profile ahead of
the photosphere in the AGB envelope. This finding is consistent with
the results of \citet{Soker1999}, who found that dust formation
occurred closer to the stellar surface ahead of magnetic-cool spots on
the equator of an AGB star. Additionally, the stellar wind critical
solution for $T_0=2000$K has a lower terminal velocity than its
counterpart for $T_0=3000$K. This is directly related to the fact that
the bulk of the gas is much cooler when the base temperature is
lower. If there exist magnetic-cool spots on the surface of an AGB
star, then the stellar wind properties, namely the wind speed and
momentum of the outflow will change, ahead of the cool spot in the
envelope of the star. This will result in asymmetric flows between
different parts of the star, in addition to the fact, that dust
formation will occur closer to the star. These will directly lead to
MHD instabilities, that can grow and become unstable and result in
asymmetric mass loss. However, it is to be noted, that investigating
such effects is beyond the scope of the current study. In order to
investigate instabilities it will be necessary to carry out MHD
calculation in at least two, if not three dimensions. By definition,
the current steady-state model cannot incorporate dynamic effects such
as instabilities. Addition of effects of radiative transfer, along with
internal chemistry, that affects dust grain formation, with relaxation
of the assumption of spherical grains, would be the ultimate goal of
such an endeavour. It has already been shown by \citet{Woitke2005}
using 2-D hydrodynamic codes with radiative transfer, that it is
possible to capture hydrodynamic instabilities. The addition of
magneto-rotational effects to these models would result in richer gas
(and dust) dynamics and quite different instabilities in the flow
altogether, due to the presence of MHD effects. The presence of
instabilities in the outflow are the precursors for asymmetric mass
loss, which has been theorised to be responsible for white dwarf kicks
\citep[see][]{Spruit1998,2008MNRAS.383L..20D,Heyl2007,Heyl2007_2,Heyl2008,Heyl2008_2,Heyl2009}. In
this context, the current work is relevant, as it presents a
steady-state solution that 2- and 3-D MHD-dust-driven wind models can
reproduce by eliminating the time dependence, i.e., setting the
$\partial/\partial t$ terms to zero. It also presents an additional avenue
for the formation of instabilities, in AGB stellar winds.

\begin{figure}
\begin{center}
\includegraphics[width=3in, scale=0.8]{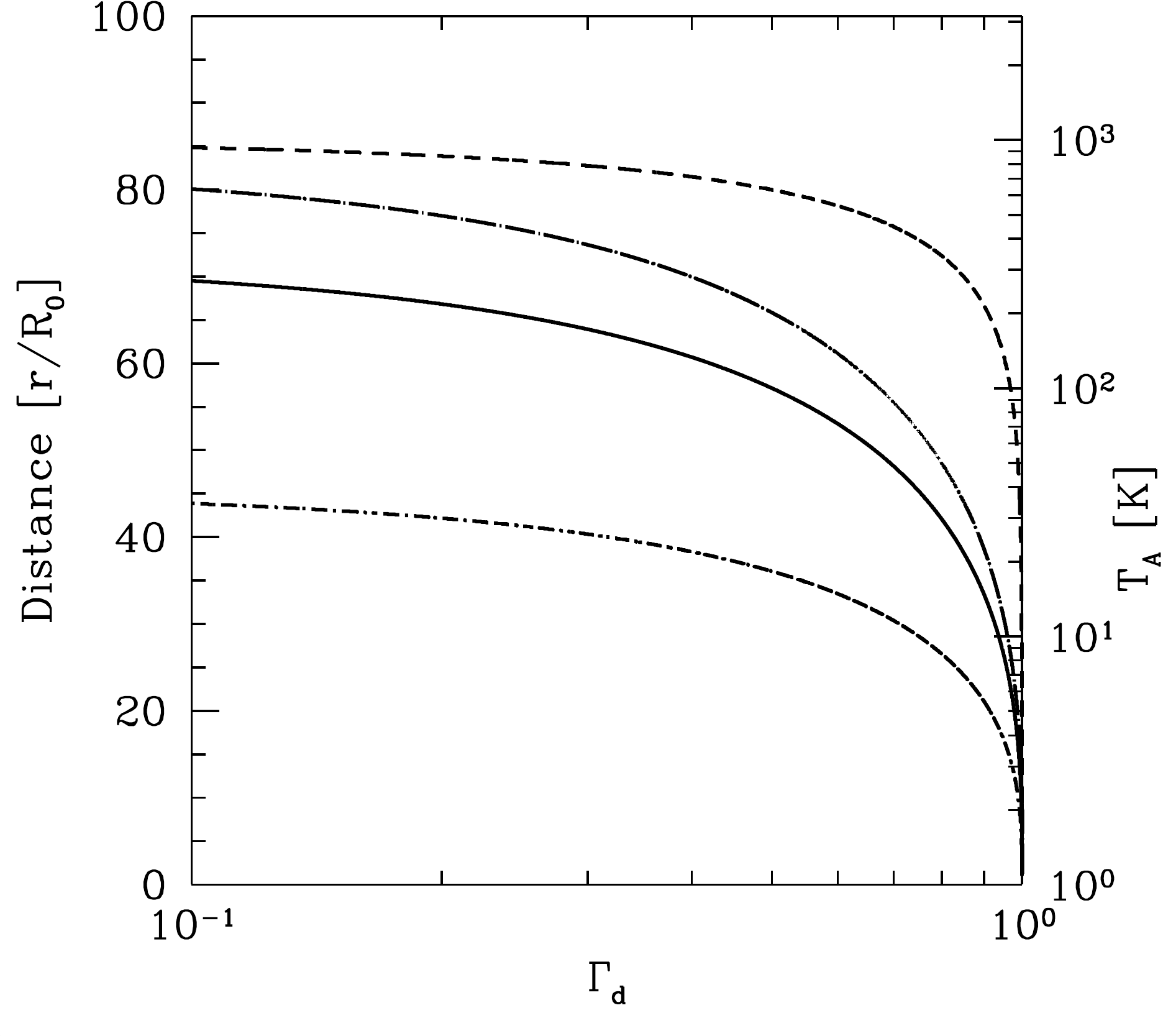}
\end{center}
\caption{Plot showing the effect of changing the dust parameter
  $\Gamma_d$ on the morphology of the family of solutions to
  Eq.~(\ref{eq:19}). In all calculations the common parameters for the
  models are shown in Table 1. The short-dash-dotted line shows
  variation in the sonic point ($r_s$) as a function of $\Gamma_d$,
  the solid line shows the change in the radial Alfv\'en point ($r_A$)
  with changing $\Gamma_d$, while the long-dash-dotted line represents
  the change in the fast point ($r_f$) for the same case. The short-dashed
  line traces the dependence of the temperature at the radial Alfv\'en
  point on $\Gamma_d$, this should be interpreted using the secondary
  axis.}
\label{fig:figure8}
\end{figure}

\begin{figure}
\begin{center}
\includegraphics[width=3in, scale=0.8]{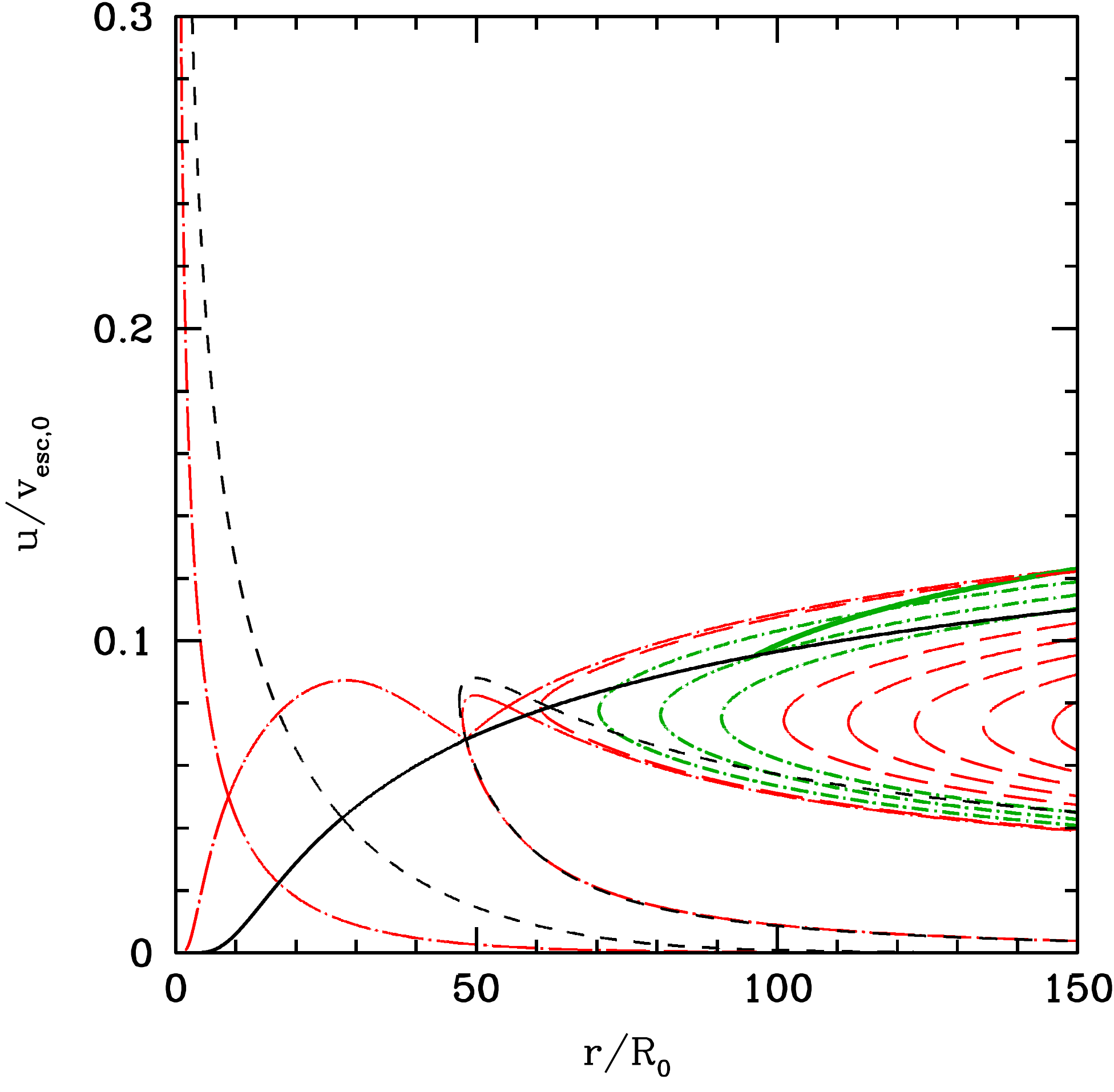}
\end{center}
\caption{Plausible hybrid wind solutions with dust formation occurring
  beyond the fast point. The red-long-dash-dotted line and the black
  solid line intersecting at the radial Alfv\'en point, are the
  critical solutions of the hybrid wind model and pure WD wind model,
  respectively. The green-short-dash-dotted lines are possible Type V
  wind solutions of a hybrid wind ($\Gamma_d = 0.3$) that can leave the star as a dust
  laden wind after dust formation occurs at the intersections with the
  solid black WD critical solution. The thick green solid line
  represents a possible hybrid wind solution with $\Gamma_d=2$.}
\label{fig:figure9}
\end{figure}

It is also to be noted that, when the dust parameter was increased to
values closer to unity, it resulted in the critical points being
suppressed towards the photosphere; this can clearly be seen in
Figure~\ref{fig:figure7}. Concordantly, as the dust parameter is
increased, it results in dust condensation closer to the stellar
surface. In Figure~\ref{fig:figure8} we have plotted the location of
the three critical points as a function of the dust parameter
$\Gamma_d$. The short-dash-dotted line represents the location of the
sonic point ($r_s$) as $\Gamma_d$ is varied, the solid line shows the
variation in the radial Alfv\'en point ($r_A$), while the
long-dash-dotted line shows the change in the fast point ($r_f$). Also
plotted therein, is the temperature at the radial Alfv\'en point; the
short-dashed line. As can be clearly seen, in the limit of $\Gamma_d
\rightarrow 1$, the three critical points converge and their locations
approach the stellar surface. The sharp decline in the
radial Alfv\'en temperature beyond $\Gamma_d \approx 0.7$ indicates
that the temperature profile in the AGB envelope falls off in an
extremely steep manner, making the solutions of the gas momentum
equation implausible and indeed unphysical.

To determine the dependence of the critical points on the dust
parameter, we continuously changed the value of $\Gamma_d$ with a step
size of $10^{-4}$, within the limits shown in Figure~\ref{fig:figure7}
and for each given value of $\Gamma_d$, we determined the appropriate
set of parameters \{$u_A$, $r_A$, $(w_s,x_s)$ and $(w_f,x_f)$\}, that
yielded a continuous monotonically increasing critical solution
through the critical points. Following which, the temperature at the
radial Alfv\'en point was determined according to Eq.~(\ref{eq:32}).

In simple isothermal dust-driven wind models, a stellar outflow is achieved by
setting $\Gamma_d > 1$ (e.g. \cite{Lamers}). This is done to
counteract the force of gravity and to drive the gas outward. However,
in the current study we found that setting the dust parameter to be
greater than unity, did not yield a set of critical points; i.e., they
were found not to exist in the domain $R_0 \leq r \leq 150R_0$, for
which Eqs.~(\ref{eq:27}-\ref{eq:30}) were satisfied
simultaneously. Hybrid winds were successfully achieved for $0 <
\Gamma_d < 1$. Indeed, as is shown in Figure~\ref{fig:figure7},
$\Gamma_d \approx 0.7$ is a reasonable physical upper limit, where
$T_A(\Gamma_d \approx 0.7) \approx 450$K and after which point the
decline in the temperature proceeds very rapidly. It is to be
mentioned that for all of the calculations carried out to produce
Figure 7, the temperature at the base of the wind was $T_0=3000$K and
all the remaining parameters were identical to those given in Table 1.

Finally, for the sake of completeness, we have shown in
Figure~\ref{fig:figure9}, the plausible hybrid wind solutions, should
the dust formation radius exist outside the fast point. Again the
parameters of the hybrid wind are identical to those shown in Table 1.
It is to be mentioned that it is likely that dust formation lies
within a few stellar radii \citep[e.g.][]{Lamers}; however, since the
parameters of Table 1 may all be scalable to suit an altogether different
type of star, we have therefore included Figure~\ref{fig:figure9}, to
complete the scenario of dust forming in the envelope or indeed the
outer atmosphere of the star.

In Figure~\ref{fig:figure9}, the solid black line represents the
critical solution of a pure Weber-Davis stellar wind, without any
dust. While the long-dash-dotted red line represents the hybrid wind
critical solution with $\Gamma_d=0.3$ and $u_A \approx 0.07v_{esc,0}$
and $r_A \approx 48.20R_0$, the pure WD wind has the same values for
$u_A$ and $r_A$. As can be seen in this case, the hybrid wind is not a
physical possibility as it is not continuous through the radial
Alfv\'en point. On the other hand the pure WD-equatorial wind is
continuous. The Bondi and Alfv\'enic solutions for the two types of
solutions, with and without dust, are also plotted in
long-dash-dotted-red and black-dashed lines respectively. The
long-dashed lines in red represent the hybrid unphysical solutions of
Type V. The intersections of the pure WD critical solution in black,
with the Type V hybrid green-short-dash-dotted lines, represent
possible locations for the dust formation radius. Thus a wind may
start off as a pure WD wind at the surface of the star, pass through
all three critical points and then undergo dust formation beyond the
fast point. At this stage the critical solution may leave the star by
following a hybrid solution represented by the short-dash-dotted lines
in green, after dust condensation. The red-long-dashed lines are not
plausible as they have either turned and become double-valued (the
left most long-dashed-red dashed line) or they do not intersect the
solid black line (the long-dashed-red lines to the right of the
green-short-dash-dotted lines), at least within the domain
indicated. The thick solid green line that intersects the solid black
line at about $r \approx 95 R_0$, is a hybrid wind solution of Type
V. This solution has the dust parameter set to a value greater than
unity ($\Gamma_d=2$). Thus it can be seen that it is possible to have
a hybrid wind with $\Gamma_d > 1$ if $r_d > r_f$. In this case, the
critical points for the hybrid wind parameters are unphysical and do not
lie within the domain $R_0 \leq r \leq 150R_0$ (see
earlier discussion relating to Figure~\ref{fig:figure8}). However in
this case, the critical wind solution has already been accelerated
through the physically possible pure-WD critical points before dust
condensation occurs in the wind. Thus, the hybrid picture can still
work with the dust parameter greater than unity as long as dust
condensation occurs beyond the fast point. As mentioned before,
Figure~\ref{fig:figure9} does not apply to AGB stars (since dust
condensation likely occurs within a few stellar radii), it is included
here for the sake of completeness and understanding the full nature of
the hybrid wind solutions.

This completes our discussion of the results of the study. In the
following section we present our summary and conclusions along side a
short discussion of directions for further investigation.

\section{Conclusion}
\label{sec:Conclusion}
We present below a brief summary of the paper and thereafter a short
discussion of possible avenues for further work.

\subsection{Summary}\label{sec:Summary}
In the preceding discussion we presented a hybrid wind model for AGB
stars. The model consists of incorporating a dust-driven wind with a
Weber-Davis MHD equatorial wind. The resulting wind momentum equations
yielded expressions for the radial and azimuthal velocities of the gas
and the dust. After eliminating the azimuthal equations, two radial
equations remained in the model that described the velocity profiles
of both the gas and the dust. In the model described in this paper, we
explicitly assumed a steady-state for the wind dynamics.

A WD wind was assumed to begin at the surface of the star, one that
would eventually fail if not for the formation of dust grains at a
given radius, which allows the hybrid wind to leave the star at
super-Alfv\'enic velocities. The dust formation was assumed to occur
abruptly, at a pre-determined radius. All the dust grains were assumed
to be perfectly spherical with identical size. It was implicitly
assumed that radiation pressure was purely in the radial direction
without scattering. The opacity of the grains was implicitly assumed
to be such that all of the radiation impinging on the grains was absorbed
and imparted momentum to the grains. The resulting drag force was
assumed to be purely radial as well. The rationale was to develop a
simple model to delineate the key points of the theory.

The hybrid wind ODE was subsequently solved using finite difference
methods, for different values of the dust parameter $\Gamma_d$. It was
found that, in order to achieve a successful hybrid wind, it was
necessary for the dust parameter to take values such that $0 <
\Gamma_d < 1$, when dust formation occurs within the slow point, i.e.,
$r_d < r_s$. The effect of changing the dust parameter revealed that
$\Gamma_d > 1$, did not yield plausible stellar wind parameters. It
was found that when $\Gamma_d \rightarrow 1$, all three critical
points converged to the the stellar surface.

Finally, the effect of changing the temperature at the base of wind
was also investigated. The temperature was changed from $T_0=3000$K to
$T_0=2000$K to represent a magnetic-cold spot on the equator of an AGB
star. It was found that lowering the temperature not only changed the
morphology of the family of solutions by suppressing the critical
points towards the stellar surface, but also required a greater value
of the dust parameter in order to achieve a successful hybrid
wind. This additionally resulted in suppressing the dust formation
radius as well, towards the photosphere of the star, consistent with
the findings of \citet{Soker1999}. Since the velocity
of the wind, ahead of the magnetic-cool spot in the AGB envelope, was
found to be appreciably lesser, than the case when the temperature was
the average equatorial temperature, it was accordingly conjectured,
that such an effect, would likely produce asymmetric outflows, owing to
the formation of MHD instabilities in the wind.

\subsection{Avenues for further investigation}\label{sec:Future work}
The question of MHD instabilities is an intriguing one. It presents a
direct route for the onset of asymmetric outflows, that have been
theorised to cause kicks to the nascent white dwarf within an AGB star
\citep[see][]{Spruit1998,2008MNRAS.383L..20D,Heyl2007,Heyl2007_2,Heyl2008,Heyl2008_2,Heyl2009}. Hydrodynamic
instabilities have already been captured in 2-D simulations of AGB
winds
\citep[see][]{Woitke2005,Woitke2005_2,Woitke2006,Woitke2008_2,Woitke2008}. Therefore
the next logical step would be to incorporate magnetic fields; a
complicated step, but one that is necessary in order to get a more
complete picture of these stars. In order to realise the onset of MHD
instabilities in the flow, a starting point would be a 2-D
axisymmetric model with magneto-rotational effects coupled with dust
formation in the envelope. The results of the current paper could be
used as a check for the steady-state solution of such a model. Such an
endeavour would undoubtedly yield interesting results and would shed
new light upon the formation of instabilities in the outflows from
these stars and answer the question of whether such instabilities can
lead to appreciably asymmetric mass loss and momentum transfer.

A second avenue, would be to relax one, or indeed several, of the
assumptions that were made in deriving the current model. As a first
experiment, it may be possible to assume that there also exists drag
in the azimuthal direction. This would result in a modification of
Eqs.~(\ref{eq:13}) and (\ref{eq:16}), to include an azimuthal drag
term. Concomitantly, the assumption that the dust-to-gas ratio is
small can be relaxed so that Eq.~(\ref{eq:4}) cannot be
approximated. This would ultimately result in four coupled ODE's (one
each for the radial and azimuthal momenta of both the gas and the
dust) that can be solved simultaneously to yield the dust and gas
velocity profiles in both the radial and azimuthal directions. Third,
the dust parameter can be assumed to vary with radius rather than be
kept fixed, this may be the easiest to implement. Depending upon the
nature of the dependence $\Gamma_d(r)$, it will change the topology of
the solution to Eq.~(\ref{eq:19}), since the solutions of
Eqs.~(\ref{eq:27}-\ref{eq:30}) will change appreciably. Fourth, the
dust grain sizes can be assumed to have a distribution, rendering the
determination of the drag force more tedious, but definitely closer to
reality. Finally, there can even be assumed to exist a certain degree
of scattering, which will effectively change the radiation pressure
term in Eq.~(\ref{eq:9}). All these changes will necessarily make the
computation more intensive, but will all ultimately yield good
dividends and further the understanding of AGB winds, complementing
the current state-of-the-art models.

\section*{Acknowledgements}
This research was supported by funding from NSERC.  The calculations
were performed on computing infrastructure purchased with funds from
the Canadian Foundation for Innovation and the British Columbia
Knowledge Development Fund. The authors would also like to thank Dr.\
Matthew Choptuick, for helpful advice regarding some of the numerical
aspects of the work. The authors are also grateful to Dr. William
T.~Vetterling, for providing some of the driver routines for
object-oriented numerical recipes in C++.

\bibliographystyle{mn2e}
\bibliography{Paper}
\label{lastpage}
\end{document}